%% file: ms.tex
\documentclass[conference]{IEEEtran}
\input{Packages.tex}
\input{Macros.tex}

\begin{document}
\input{0_Titlepage.tex}

\input{1_Introduction.tex}
\input{2_SystemModel.tex}

\input{3_LocalizationSchemes.tex}
\input{4_Bounds.tex}
\input{5_Evaluation.tex}
\input{6_Conclusion.tex}
\input{AppendixA.tex}

\balance

\bibliographystyle{IEEEtran}
\bibliography{ref.bib}

\end{document}

%% file: Packages.tex
\usepackage{cite}
\usepackage{amsmath,amssymb,amsfonts}
\usepackage{algorithmic}
\usepackage{graphicx}
\usepackage{textcomp}
\usepackage{tikz,pgfplots,pgfplotstable}
\usepackage{multicol}
\usepackage{multirow}
\usepackage[american, EFvoltages]{circuitikz}
\usepackage{stringstrings}
\usepackage{adjustbox}
\usepackage{siunitx}
\usepackage{paralist}
\usepackage[font=footnotesize]{caption}
\usepackage{subcaption}
\usepackage{bm}
\usepackage{balance}

\usetikzlibrary{arrows.meta}
\usetikzlibrary{calc}
\usetikzlibrary{backgrounds}

\pgfplotsset{compat=newest}
\pgfplotsset{every axis/.append style={
                    label style={font=\small},
                    tick label style={font=\small}  
                    }}

\def\BibTeX{{\rm B\kern-.05em{\sc i\kern-.025em b}\kern-.08em
    T\kern-.1667em\lower.7ex\hbox{E}\kern-.125emX}}
    \newcommand{\norm}[1]{\ensuremath{\left \lVert #1 \right \rVert}}    

   \makeatletter
\renewcommand*\env@matrix[1][*\c@MaxMatrixCols c]{%
  \hskip -\arraycolsep
  \let\@ifnextchar\new@ifnextchar
  \array{#1}}
\makeatother

 \usepackage{diagbox}
\usepackage{siunitx}
\usepackage[mode=image]{standalone}
\usepackage{makecell} 
\setcellgapes{5pt}

\usepackage{cleveref}
\crefname{figure}{Fig.}{Fig.}
\Crefname{figure}{Fig.}{Fig.}
\crefname{section}{Sec.}{Sec.}
\Crefname{section}{Sec.}{Sec.}
\crefname{appendix}{App.}{App.}
\Crefname{appendix}{App.}{App.}
\crefname{subsection}{Sec.}{Sec.}
\Crefname{subsection}{Sec.}{Sec.}
\crefname{theorem}{Theorem}{Theorem}
\Crefname{theorem}{Theorem}{Theorem}
\crefname{lemma}{Lemma}{Lemma}
\Crefname{lemma}{Lemma}{Lemma}
\crefname{prop}{Prop.}{Prop.} 
\Crefname{prop}{Prop.}{Prop.}
\crefname{equation}{}{}
\Crefname{equation}{}{}

%% file: Macros.tex
\newcommand\tn[1]{\text{\normalfont{#1}}} 

\newcommand{\FIM}{\bm{\mathcal{I}}}

\newcommand{\PSI}{\boldsymbol{\psi}}
\newcommand{\gpsi}[2]{\frac{\partial {#1}}{\partial #2}}

 \definecolor{matBlue}{rgb}		{0, 0.447, 0.741}
\definecolor{matGreen}{rgb}		{0.466, 0.674, 0.188}
\definecolor{matOpacGreen}{rgb}		{0.843, 0.902, 0.765			}
\definecolor{matRed}{rgb}		{0.85, 0.325, 0.098	}
\definecolor{matYellow}{rgb}		{0.929, 0.694, 0.125			}
\definecolor{matOpacYellow}{rgb}		{0.976, 0.910, 0.761			}
\definecolor{matPurple}{rgb}		{0.494, 0.184, 0.556	}
\definecolor{matLblue}{rgb}		{0.301, 0.745, 0.933	}
\definecolor{matBrown}{rgb}		{0.635 ,0.078 ,0.184	}
\definecolor{matBlack}{rgb}		{0,				0,				0				}
\definecolor{matGray}{rgb}		{0.5,				0.5,				0.5				}
 \definecolor{dpebRed}{rgb}{0.4, 0.4, 0.4}
 \definecolor{gridGray}{rgb}{ 0.89, 0.89, 0.89} 

  \newcommand\f[2]{\frac{#1}{#2}} 
  \newcommand\fp[2]{\f{\partial #1}{\partial #2}} 
  \newcommand\re{\mathrm{Re}} 
  \newcommand\im{\mathrm{Im}} 
\renewcommand\Re{\re} 
\renewcommand\Im{\im} 
\DeclareMathOperator*\argmax{arg\,max}
\DeclareMathOperator*\argmin{arg\,min}

  \newcommand\mtx[2]{\left[\begin{array}{#1}#2\end{array}\right]}
  \newcommand\Tr{^\mathrm{\normalfont{{T}}}}     
\newcommand\Hr{^\mathrm{\normalfont{H}}}        
\DeclareMathOperator*\diag{diag}   
\DeclareMathOperator\trace{tr} 

  \newcommand\p{\mathbf{p}}   
\renewcommand\o{\mathbf{o}}   
 
\newcommand\bH{\mathbf{H}}  
\newcommand\bHN{\breve{\mathbf{H}}}  
\newcommand\PEB{\mathrm{PEB}}

\newcommand\meass[1]{^{\tn{meas}#1}}

\newcommand\meas{\meass{}}

\newcommand\eqskip{\, ,}
\newcommand\eqend{\, .}
\newcommand\inv{^{-1}}

\newcommand{\n}{n}
\newcommand{\m}{m}
\newcommand{\N}{N}
\newcommand{\M}{M}
\newcommand{\cpair}{\ensuremath{\m, \n}}
\newcommand{\csub}{\ensuremath{_{\cpair}}}

\newcommand{\ncoop}{\ensuremath{^{\mathrm{NonCoop}}}}
\newcommand{\coop}{\ensuremath{^{\mathrm{Coop}}}}

\renewcommand{\norm}[1]{\lVert #1 \rVert}
\newcommand{\F}{\ensuremath{\mathbf{F}\csub}}
\renewcommand{\O}{\ensuremath{\mathbf{O}}}
\newcommand{\TA}{three-axis }
\newcommand{\TAC}{Three-Axis }
\newcommand{\NC}{non-cooperative }
\newcommand{\numLS}{numLS}
\newcommand{\pairML}{pairML}
\newcommand{\turboLS}{turboLS}
\newcommand{\coeff}{a} 

\newcommand\mylinewidth{0.75pt}

\newcommand{\plotPEB}[1]{\addplot [color=#1, line width=\mylinewidth, mark=triangle, mark size=4.8pt, mark options={solid, #1}]}
\newcommand{\plotPinit}[2]{\addplot [color=#1, line width=\mylinewidth, mark=x, mark size=3.2pt, mark options={solid, #1}, #2]}
\newcommand{\plotAdvinit}[2]{\addplot [color=#1, line width=\mylinewidth, mark=square, mark size=2.5pt, mark options={solid, #1}, #2]}
\newcommand{\plotML}[2]{\addplot [color=#1, line width=\mylinewidth, mark=o, mark size=2.5pt, mark options={solid, #1}, #2]}
\newcommand{\plotMultilat}[2]{\addplot [color=#1, line width=\mylinewidth, mark=diamond, mark size=3.0pt, mark options={solid, #1}, #2]}
\newcommand{\plotRinitOne}[2]{\addplot [color=#1, line width=\mylinewidth, mark=asterisk, mark size=2.9pt, mark options={solid, #1}, #2]}
\newcommand{\plotRinitFive}[2]{\addplot [color=#1, line width=\mylinewidth, mark=pentagon, mark size=2.7pt, mark options={solid, #1}, #2]}


%% file: 0_Titlepage.tex
\pgfplotsset{try min ticks=3}
\title{Cooperative Magneto-Inductive Localization}
\author{\IEEEauthorblockN{Henry Schulten, Gregor Dumphart, Antonios Koskinas and Armin Wittneben}
\IEEEauthorblockA{Wireless Communications Group, D-ITET, ETH Zurich, Switzerland \\ 
E-Mail: \{schulten, dumphart, koskinas, wittneben\}@nari.ee.ethz.ch} 
}

\maketitle
\begin{abstract}
Wireless localization is a key requirement for many applications. It concerns position estimation of mobile nodes (agents) relative to fixed nodes (anchors) from wireless channel measurements. Cooperative localization is an advanced concept that considers the joint estimation of multiple agent positions based on channel measurements of all agent-anchor links together with all agent-agent links.
In this paper we present the first study of cooperative localization for magneto-inductive wireless sensor networks, which are of technological interest due to good material penetration and channel predictability. We demonstrate significant accuracy improvements (a factor of $3$ for $10$ cooperating agents) over the \NC scheme.
The evaluation uses the Cram\'er-Rao lower bound on the cooperative position estimation error, which is derived herein.
To realize this accuracy, the maximum-likelihood estimate (MLE) must be computed by solving a high-dimensional least-squares problem, whereby convergence to local minima proves to be problematic. A proposed cooperative localization algorithm addresses this issue: first, preliminary estimates of the agent positions and orientations are computed, which then serve as initial values for a gradient search. In all our test cases, this method yields the MLE and the associated high accuracy (comprising the cooperation gain) from a single solver run. The preliminary estimates use novel closed-form MLE formulas of the distance, direction and orientation for single links between \TA coils, which are given in detail.
\end{abstract}


%% file: 1_Introduction.tex
\section{Introduction}
\label{sec:Intro}

 Low-frequency magnetic induction as a transmission mechanism promises strong links through resonant multi-turn coil designs and good material penetration because of vastly limited shadowing and multipath fading. It is therefore an interesting propagation mechanism for various applications in wireless sensor networks (WSN) or the Internet of Things (IoT), e.g. for underground \cite{bib:abrudan2015distortion}, underwater \cite{bib:akyildiz2015water}, and biomedical \cite{bib:Sitti2015} environments. A key requirement for many such applications is wireless localization, i.e. the ability to compute an accurate estimate of a wireless node's position and possibly its orientation from wireless channel measurements.

In many cases a magneto-inductive channel can be described with a simple free-space dipole model in good approximation. This predictable behavior makes magnetic induction well-suited for wireless localization by parameter estimation \cite{bib:dumphart2020magneto}, i.e. by computing the location parameters which yield the best fit between channel model and noisy measurements (e.g., by solving a least-squares problem). This approach has been studied for an agent node equipped with a single coil using a single temporal snapshot of channel measurements in \cite{bib:Schlageter2001,bib:SongHu2009,bib:dumphart2017robust,bib:Hu2005} and for a \TA coil using consecutive channel measurements (tracking) in \cite{bib:abrudan2015distortion}. A prominent alternative approach is magnetic location fingerprinting, which relies on a prerecorded database of locations and associated measurements \cite{bib:XieTMC2015}.

Magneto-inductive localization via parameter estimation suffers from local minima of the cost function: when a numerical solver converges to a local minimum, then the estimation error will usually be large \cite{bib:dumphart2017robust,bib:Hu2005}. Addressing this robustness issue with a global optimization approach (e.g., multistart gradient search) requires long computation times which may prohibit real-time use. Furthermore, practical magneto-inductive localization systems have limited accuracy whenever the assumed channel model has limited applicability, e.g. when unconsidered conducting objects are nearby and cause field distortions \cite{bib:KyprisTGRS2016,bib:dumphart2019practical}.

We consider a scenario with multiple mobile nodes (agents). Conventionally, one would carry out parameter estimation for each agent individually, based on the channel measurements between this agent and the fixed infrastructure nodes (anchors) \cite{bib:Schlageter2001,bib:SongHu2009,bib:dumphart2017robust,bib:Hu2005}. Cooperative localization instead considers joint parameter estimation for all agents, based on all agent-anchor measurements together with all agent-agent measurements. The idea originates from the ultra-wideband localization literature \cite{bib:shen2010fundamental,bib:WinCM2011,bib:buehrer2018collaborative}, wherein accuracy and robustness improvements were identified as benefits of cooperation. They arise from the effect of agent-agent measurements in joint estimation \cite{bib:WinCM2011}.

This paper applies the cooperative localization idea to magneto-inductive localization, with the goal of alleviating the described accuracy and robustness problems. In particular, we make the following contributions:
\begin{itemize}
\item For a system model which assumes a \TA coil at each agent and each anchor, we state the least-square estimate (LSE) in general form, for cooperative and \NC localization.
\item For the case of Gaussian-distributed measurement errors we state the Fisher information matrix for the unknown position and orientation parameters of multiple agents. We use the resultant Cram\'er-Rao lower bound on the cooperative position estimation error to demonstrate the accuracy improvement obtained through cooperation.
\item We provide a closed-form expression for the joint MLE of distance, direction, and orientation based on measurements of the $3 \times 3$ channel matrix between a pair of \TA nodes. We also show how these can be used to obtain an accurate position estimate from the observations of a single link.
\item We propose a localization algorithm named turboLS which uses the aforementioned MLE to initialize the LSE computation (a Levenberg-Marquardt search) for cooperative networks. We demonstrate an empirical robustness of $100\%$ through numerical evaluation.
\end{itemize}

\subsubsection*{Paper Structure}
We present the employed system model in \Cref{sec:systemmodel}, the considered localization algorithms in \Cref{sec:localization} and the Fisher information analysis in \Cref{sec:bounds}. The numerical results follow in \Cref{sec:eval} and the conclusions in \Cref{sec:conclusion}.

%% file: 2_SystemModel.tex
\section{System Model}
\label{sec:systemmodel}

\subsection{Setup Description}
\begin{figure}[!htb]
\centering
\input{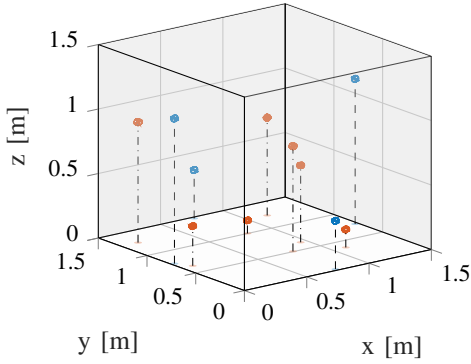}
\caption{Three-dimensional cubic setup with $\M=7$ agents (red) that are to be localized and $\N=4$ anchors (blue).}
\label{fig:3DRoom}
\end{figure}

In this paper, we will look at an IoT-motivated setup which comprises a $\SI{1.5}{\meter}$ cubic volume (e.g. a small office room) with fixed anchors on its lateral boundaries (e.g. the room walls) and multiple randomly placed agents within the room.  The deployment of the anchors is assumed to be fully known, whereas the deployment of the agents is unknown. Each of these sensor nodes is a so called \TA coil consisting of three orthogonal solenoid coils. The goal of this work is to estimate all agent positions and thereby reconstruct the entire sensor network.  The illustration in \Cref{fig:3DRoom} shows an exemplary setup with $\M = 7$ agents (red) and $\N =4$ anchors (blue).

\subsection{Channel Gains between Weakly Coupled \TAC Coils}
As was shown in \cite{bib:dumphart2017robust,bib:dumphart2020magneto}, the localization of single solenoid coils is often problematic due to the strong orientation dependency of their channel gains. While cooperation might mitigate these orientation issues to some extent, we exclusively consider nodes with \TA coils in order to ensure reasonably strong links between any node pair.

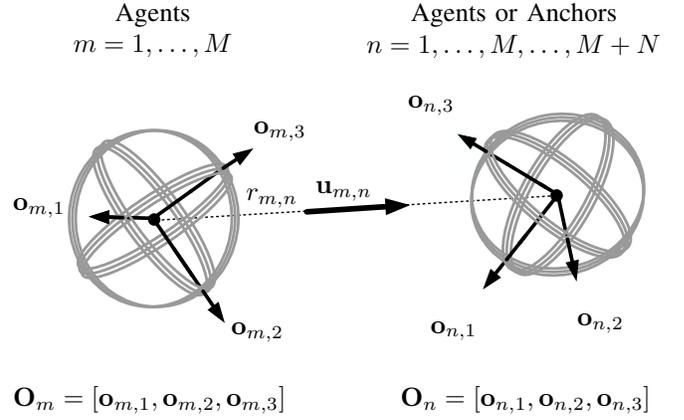
\begin{figure}
\centering
\input{TriAxialPair_MIMO_gray}
\caption{A pair of magneto-inductive nodes $(\cpair)$ with \TA coils, comprising a transmitting agent $m$ and a receiving node $n$ that is either an agent or anchor. Each \TA coil consists of three orthogonal solenoid subcoils. The node orientations $\O_\m , \O_\n \in \mathbb{R}^{3 \times 3}$ are arbitrary orthogonal matrices. The figure and formalism are adopted from \cite{bib:dumphart2016stochastic,bib:dumphart2020magneto}.}
\label{fig:3AxCoils}
\end{figure}

\Cref{fig:3AxCoils} illustrates a pair (\cpair) of such \TA coils in an arbitrary arrangement.  Thereby,  the index $m=1, \hdots, \M$ is used to exclusively describe the agents, while the index $n=1, \hdots, \M, \hdots, \M+\N$ is used to describe both agents and anchors. The deployment of any coil $n$ is then defined by its three-dimensional central position $\mathbf{p}_\n$ and the corresponding Cartesian unit orientation vectors of each single subcoil, which can be summarized as an orientation matrix $\O_\n=[\mathbf{o}_{\n,1}, \mathbf{o}_{\n,2}, \mathbf{o}_{\n,3}]$. The  distance vector of the shown coil pair is given by $\mathbf{r}\csub=\mathbf{p}_\n - \mathbf{p}_\m $ with distance $r\csub= \norm{\mathbf{r}\csub }$ (Euclidean norm) and unit direction $\mathbf{u}\csub=\f{\mathbf{r}\csub}{r\csub}$.

The complex-valued channel matrix $\bH\csub \in \mathbb{C}^{3 \times 3}$ between a node pair $(m,n)$ is given by \cite{bib:dumphart2016stochastic,bib:abrudan2015distortion,bib:dumphart2020magneto}
\begin{align}
\bH\csub = \frac{j\coeff\csub}{r\csub^3} \; \; \O_\n \Tr \left(\frac{3}{2}\mathbf{u}\csub\mathbf{u}\csub\Tr - \frac{1}{2} \, \mathbf{I}_3 \right) \O_\m \eqskip  
\label{eq:system_pairwise_channel_matrix}
\end{align}
where $\mathbf{I}_3$ is the $3 \times 3$ identity matrix and
\begin{align}
\coeff \csub = \f{\mu \, A_\n A_\m \, \nu_\n \nu_\m}{\sqrt{4 R_\n R_\m}} f \eqskip
\end{align}
is a technical parameter of the involved subcoils.
Here, $ A$ describes the subcoil's surface area,  $\nu$ the number of coil windings and $R$ the ohmic losses. Moreover, $f$ is the chosen operating frequency and $\mu$ is the  magnetic permeability.

All measured channel gains are subject to additional perturbations due to receiver noise, interference, environmental clutter, material tolerances and other real-world effects. In \cite{bib:dumphart2019practical} it was shown that these impacts may be approximated as additive errors $\mathbf{W}\csub$ on the channel gains
\begin{align}
&\bH\meas\csub = \bH\csub+ \mathbf{W}\csub \eqskip
\label{eq:system_observation_vector} 
\end{align}
with a circularly-symmetric complex Gaussian distribution
$[\mathbf{W}\csub]_{k,l} \overset{\mathrm{i.i.d}}{\sim} \mathcal{CN}(0,\sigma^2)$. Thereby $k,l \in \{1,2,3\}$ are matrix element indices and $\sigma^2$ is the measurement error variance.

For our proposed application, a multitude of these channel matrix measurements are acquired and subsequently processed in order to estimate all agent positions. We consider two different scenarios:
\begin{itemize}
    \item \textbf{Non-Cooperative Scenario:} We consider those channel matrices $\bH\csub\meas$  that correspond to agent-anchor links between an agent node $\m \in \{1, \hdots, \M \}$ and an anchor node $\n \in \{ \M+1, \hdots, \M+\N \}$.
    \item \textbf{Cooperative Scenario:} We additionally consider the agent-agent links. In total, we consider $\bH\csub\meas$ for all links between an agent node $\m \in \{1, \hdots, \M \}$ and any other node (agent or anchor) $\n \in \{ 1, \hdots, \M+\N \} \setminus \{ \m \}$.
\end{itemize}

%% file: TriAxialPair_MIMO_gray.tex
%
%
\begin{tikzpicture}
\node[draw=none,fill=none] at (0,0){  \includegraphics[width=0.84\columnwidth,trim=3cm 5cm 5cm 7cm]{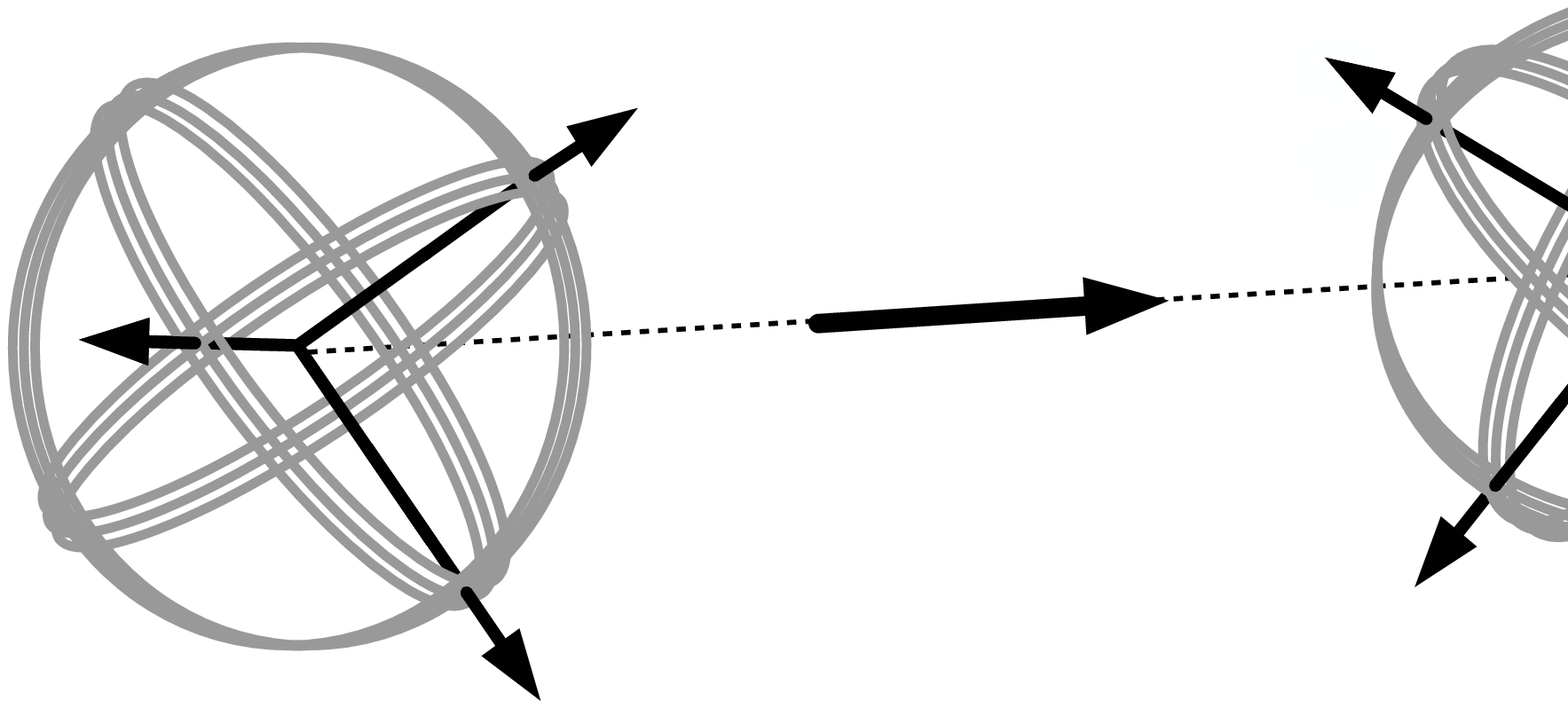}};

\node[circle, fill=black, scale=0.5] (M) at (-2.32,0.14) {};
\node[circle, fill=black, scale=0.5] (N) at (3.03,0.46) {};
\draw coordinate (A) at (-0.3,0.52);

\node[right, align=left] at (A) {$\mathbf{u}\csub$};
\node[right, align=left] at ($(A) + (-0.92, -0.1)$) {$r\csub$};

\node[right, align=left] at ($(M) + (-2, 0.15)$) {$\o_{\m,1}$};
\node[right, align=left] at ($(M) + (0.9, -1.5)$) {$\o_{\m,2}$};
\node[right, align=left] at ($(M) + (1.2, 1.175)$)  {$\o_{\m,3}$};

\node[right, align=left] at ($(N) + (-1.8,-1.8)$) {$\o_{\n,1}$};
\node[right, align=left] at ($(N) + (0.15,-1.7)$) {$\o_{\n,2}$};
\node[right, align=left] at ($(N) + (-2.1,1.2)$) {$\o_{\n,3}$};

\node[right, align=left] at ($(M) + (-2,-2.4)$) {$\O_{\m}=[\o_{\m,1}, \o_{\m,2}, \o_{\m,3}]$};
\node[right, align=left] at ($(N) + (-2.2,-2.72)$) {$\O_{\n}=[\o_{\n,1}, \o_{\n,2}, \o_{\n,3}]$};

\node[right, align=center] at ($(M) + (-1.2,2.52)$)  {Agents \\ $\m=1, \hdots, \M$};
\node[right, align=center] at ($(N) + (-2.65,2.2)$) {Agents or Anchors \\ $\n=1, \hdots, \M, \hdots, \M + \N$};

\end{tikzpicture}%
%
%
%
%
%

%% file: 3_LocalizationSchemes.tex
\section{Localization Schemes}
\label{sec:localization}
As described in \Cref{sec:systemmodel}, each agent coil's deployment is fully described by its three-dimensional position $\p_\m$ and its orientation matrix $\O_\m$, which in turn is fully defined by three Euler angles $\boldsymbol{\phi}_\m=[\alpha_\m, \beta_\m, \gamma_\m]\Tr$ or their corresponding rotation matrices. For each agent coil, we thus have to estimate these six parameters%
\footnote{Even when the focus lies on estimating the position parameters, the orientation parameters have to be included in the estimation problem as nuisance parameters.}
that we summarize as deployment vector $\boldsymbol{\psi}_\m=[\p_\m\Tr, \boldsymbol{\phi}_\m\Tr]\Tr$.  In the following section, we will provide three different approaches to achieve this goal, which are based on (\textit{A}) a numerical least-squares  (LS) estimation, (\textit{B}) a pairwise maximum likelihood (ML) distance estimation with subsequent position estimation, and (\textit{C}) a combination of both. All approaches can be performed for either cooperative  or \NC channel matrix observations.

\subsection{Numerical Least-Squares Estimation (\numLS)}
\label{subsec:LS}
The availability of the statistical model for the measured channel gains in \eqref{eq:system_observation_vector} allows for a straightforward formulation of the joint MLE of the unknown agent deployments. In our case the MLE coincides with the LSE because of the zero-mean i.i.d. Gaussian errors. For the cooperative scenario, this estimator solves the $6M$-dimensional nonlinear problem
\begin{align}
(\hat{\PSI}_1, \hdots, \hat{\PSI}_\M)\coop
=
\argmin_{{\PSI}_1, \hdots, {\PSI}_\M}
\sum _{m=1}^{\M}  \sum _{\substack{\n=1 \\ \n \neq \m}}^{\M+\N}
\big\Vert \bH\meas\csub-\bH\csub \big\Vert^2 _\mathrm{F}
\label{eq:localization_LS_coop}
\end{align}
where $\bH\csub$ is a function of $\PSI_\m$ and $\PSI_\n$ if $\n \in \{1,\ldots,\M\}$ (agent-agent link) or a function of just $\PSI_\m$ if $\n > \M$ (agent-anchor link with known anchor parameters).

For the \NC scenario, the problem decomposes into $\M$ individual six-dimensional nonlinear optimization problems (one per agent), 
\begin{align}
\hat{\PSI}_\m\ncoop
=
\argmin_{{\PSI}_\m} \sum _{n=\M+1}^{\M+\N}
\big\Vert \bH\meas\csub-\bH\csub(\PSI _\m) \big\Vert^2 _\mathrm{F}\eqend
\label{eq:localization_LS_noncoop}
\end{align}
To the best of our knowledge, \Cref{eq:localization_LS_coop} and \Cref{eq:localization_LS_noncoop} can only be solved via numerical methods such as an iterative gradient-based search. For the optimization problems at hand, a gradient search however turns out to be computationally expensive and very sensitive to the provided initial value: it is prone to converge to local minima, especially in the high-dimensional cooperative scenario. These numerical challenges may practically nullify the accuracy-improvement potential of cooperative localization, unless they are alleviated by providing a reasonably accurate initial value.

\subsection{Pairwise Estimation (\pairML)}
\label{subsec:pairwise_ML}

In order to alleviate the problems described in \Cref{subsec:LS} we propose a second approach, which is based on the individual ML estimation for the pairwise links between agents and anchors. We consider the log-likelihood function (log-LHF) $ L( \PSI _\m)$ of an agent's deployment vector based on one measured pairwise agent-anchor channel matrix $\bH\meas \csub$  (with $\n>\M$). For mathematical convenience we discard constant summands from $L( \PSI _\m)$ and define $\F  = \left(\frac{3}{2}\mathbf{u}\csub\mathbf{u}\csub\Tr - \frac{1}{2}\mathbf{I}_3 \right)$ and $ \bHN \csub   =  \frac{j\coeff\csub}{r\csub^3} \; \F$ as in \cite{bib:dumphart2020magneto}.
We find that
\begin{align*}
    L( \PSI _\m) &= -\frac{1}{\sigma^2}\trace\left ( \left(\bH\meas \csub - \bH \csub \right)\Hr \left (\bH\meas \csub - \bH \csub \right) \right ) \\
    &\propto -\trace\left ( \left(\bH\meas \csub \right)\Hr \bH\meas \csub \right ) - \trace\left ( \bHN\csub\Hr \bHN\csub \right )  \\
    &\hphantom{\propto ,}+ 2\Re \left \{\trace\left (\bH\meas \csub)\Hr \O_\n \Tr \bHN \csub \O_m \right ) \right \}
    \\
    &= -\trace\left ( \left(\bH\meas \csub \right)\Hr \bH\meas \csub \right ) - \frac{3}{2} \cdot \frac{\coeff^2\csub}{r\csub ^6}  \\
    &\hphantom{= ,}+   \frac{2 \coeff \csub}{r\csub^3} \trace\Big ( \underbrace{\Im\!\left \{ \bH\meas \csub\right \}\Tr \O_\n\Tr}_{=:\,\mathbf{A}\csub}  \; \underbrace{\F \O_m}_{=:\, \mathbf{B}\csub} \Big )
\end{align*}
where the first summand is independent of $\PSI _\m$. The maximization of the log-LHF thus simplifies to
\begin{align}
   \hat{\PSI}_\m^\mathrm{ML} = \argmax_{\PSI _\m}  - \frac{3 \coeff^2 \csub}{2 r\csub ^6}  +  \frac{2 \coeff \csub}{r\csub^3} \trace\left ( \mathbf{A}\csub \mathbf{B}\csub \right ) \eqskip
\end{align}
where $\mathbf{A}\csub \overset{\mathrm{SVD}}{=}\mathbf{U}\csub\mathbf{S}\csub \mathbf{V}\csub\Tr$ is fixed based on the measurement matrix as well as the anchor orientation, and $\mathbf{B}\csub$ is the optimization variable under the structural constraints of both $\F$ and $\O_\m$. 

We consequently need to maximize $\trace (\mathbf{A}\csub\mathbf{B}\csub)$ with respect to $\F$ and $\O_\m$ but independently of the distance. According to Von Neumann's trace inequality (cf. \cite{bib:mirsky1975traceNeumann}), this term is generally maximized if both matrices share the same singular vectors. However, this may not be feasible when $\O_\m$ is required to be a proper rotation matrix. Our problem hence  becomes a specific constrained form of the Procrustes problem (cf.  \cite{bib:schonemann1966generalizedProcrustes}) and is closely related to the Kabsch algorithm (cf. \cite{bib:kabsch1976solution}). Its solution is obtained via 
\begin{align}
& \hat{\O}_m ^\mathrm{ML}  = \mathbf{V}\csub  \diag \left (1, -1, -\det \left ( \mathbf{U}\csub\mathbf{V}\csub\Tr \right) \right )
\mathbf{U}\csub\Tr
\eqskip \label{eq:pairwiseMLorientation}\\ &
\hat{\mathbf{F}}\csub ^\mathrm{ML} = \mathbf{V}\csub \diag \left (1, -\frac{1}{2}, -\frac{1}{2} \right ) \mathbf{V}\csub\Tr
\end{align}
and hence the maximum value of the trace is equal to 
${z\csub=\trace \Big (\mathbf{S}\csub \cdot \diag \left (1, \frac{1}{2}, \frac{1}{2} \cdot \det( \mathbf{U}\csub\mathbf{V}\csub\Tr ) \right )} \Big )$.
For any given distance $r\csub$, this solution maximizes the likelihood function while satisfying the structural constraints of both estimated matrices. The associated ML distance estimate is then found by solving
$\vphantom{\bigg (}-\frac{9 \coeff^2\csub}{r\csub ^7}  + \frac{6 \coeff \csub z\csub}{r\csub^4}  \overset{!}{=} 0$,
yielding
\begin{align}
\hat{r}\csub^\mathrm{ML} =\sqrt[\text{\normalsize $3$}]{\ \frac{3}{2} \cdot \frac{\coeff \csub}{z\csub}}
\label{eq:pairwiseMLdistance}
\end{align}
where the real-valued root is chosen as estimate.

The above solutions have the following essential consequence: the estimate $\hat{\mathbf{F}}\csub ^\mathrm{ML}$ entails an estimate $\hat{\mathbf{u}}\csub$ of the unit direction vector between agent and anchor, up to an ambiguity of its sign. It is given by the singular vector of $\mathbf{V}\csub$ that corresponds to the largest singular value. Similar to \cite{bib:abrudan2015distortion}, we can hence obtain an estimate on the agent position as
\begin{align}
    \hat{\mathbf{p}}_\m = \mathbf{p}_\n \pm \hat{\mathbf{u}}\csub \, \hat{r}\csub^\mathrm{ML} \eqend
    \label{eq:pairwiseMLposition}
\end{align}
In our described setup, we choose to resolve the ambiguity by just deciding on the solution that is within the room. In combination with the orientation estimate $\hat{\O}_m ^\mathrm{ML}$, which can be transformed to an estimate of $\boldsymbol{\phi}_\m$ by the corresponding trigonometric relationships, we can ultimately estimate the full agent deployment $\PSI _\m$ based on each single measurement of the $3 \times 3$ channel matrix between an agent and an anchor.

\subsection{Advanced LS Initialization (\turboLS)}
\label{subsec:advancedInit}
The previously introduced estimation approach can either be used as a position estimator in its own right or the obtained estimates may serve as initial values for the vast numerical optimization problem associated with LS estimation. Having an improved initial value close to the global minimum of the high-dimensional cost function can substantially reduce the computation time and  the chance of missing the global minimum. We will hence also consider an estimation approach where both the position and orientation estimates of \Cref{subsec:pairwise_ML} are used as initialization values for the cooperative and \NC LS estimators of \Cref{subsec:LS}. However, as we obtain multiple potential initialization points due to having multiple anchors per agent, we will only choose the estimates of the pair with the smallest distance estimate (which will most likely be the link with the best signal-to-noise ratio). Should this pair not have a single valid position estimate (i.e. both position estimates are either outside or within the room), then we iteratively resort to the pair which has the next smallest distance estimate.

%% file: 4_Bounds.tex
\section{Position Error Bounds}
\label{sec:bounds}

In order to compare the accuracy of the previously introduced LS estimators to a meaningful theoretical limit, this section derives the Position Error Bound (PEB), which is the Cram\'er-Rao Lower Bound (CRLB) on the position root-mean-square error (RMSE) that applies to any unbiased estimator. To this end, we first need to compute the corresponding Fisher Information Matrices (FIMs). 
For the cooperative scheme that considers all inter-node observations, the $6M \times 6M$ FIM is analogous to \cite[Eq.~(12)]{bib:shen2010fundamental} and has a $6 \times 6$ block structure 
\begin{align} 
\FIM \coop =\! \mtx{cccc}{
\! \mathbf{N}_1 + \mathbf{C}_{1,1} \!\!\! &\mathbf{C}_{1,2} & \cdots & \mathbf{C}_{1,\M} \\
\mathbf{C}_{2,1} & \!\!\! \mathbf{N}_2 + \mathbf{C}_{2,2} \!\!\!\!\! & \cdots & \mathbf{C}_{2,\M} \\
\vdots & \vdots & \ddots & \vdots \\
\mathbf{C}_{\M,1 }& \mathbf{C}_{\M,2 } & \cdots & \!\!\! \mathbf{N}_\M + \mathbf{C}_{\M,\M} \!\!\! }
\label{eq:FIM_coop}
\end{align}
where the scalar elements of the individual submatrices are given by
\begin{align}
&
[\mathbf{N}_\m]_{k,l} = \f{2}{\sigma ^2}  \sum _{{\n=M+1}}^{\M+\N}  \trace\!\left ( \fp{\bH \Hr  \csub }{[\PSI _\m ]_k} \fp{\bH  \csub }{[\PSI _\m ]_l}  \right)
\eqskip \\ &
[\mathbf{C}_{\m , \m}]_{k,l} = \f{2}{\sigma ^2}  \sum _{\substack{\n=1 \\ \n \neq \m}}^{\M}   \trace\!\left ( \fp{\bH \Hr  \csub }{[\PSI _\m ]_k} \fp{\bH  \csub }{[\PSI _\m ]_l}  \right)
\eqskip \\ &
[\mathbf{C}_{\m , \n}]_{k,l} = \left. \f{2}{\sigma ^2}    \trace \!\left ( \fp{\bH \Hr  \csub }{[\PSI _\m ]_k} \fp{\bH  \csub }{[\PSI _\n ]_l}  \right) \right\vert_{n\in \{1 \hdots \M \} \setminus \{\m\}}
\label{eq:FIM_noncoop}
\end{align}
with $k,l \in \{ 1, \hdots, 6 \}$. The required derivatives of the channel matrices are given in Appendix \ref{app:A_deriv}. The diagonal blocks $\mathbf{N}_\m$ account for the measurements from agent $\m$ to all anchors, whereas the diagonal blocks $\mathbf{C}_{\m , \m}$ account for all inter-agent measurements of this agent, in case the other agents act as anchors. However, as their deployment is unknown, this uncertainty in their function as anchor is expressed by the off-diagonal blocks $\mathbf{C}_{\m , \n}$.

Consequently, in the cooperative case, the PEB of agent $m$ is found as
\begin{align}
\PEB_m \coop = \sqrt{ \sum_{k=1+6(m-1)}^{3+6(m-1)} \Big[\big(\FIM\coop \big)\inv \Big ]_{k,k} } \eqend
\end{align}

In the \NC case, the absence of inter-agent measurements causes $\mathbf{C}_{\m , \m} = \mathbf{0}$ and $\mathbf{C}\csub = \mathbf{0}$ $\forall m,n$. The FIM \eqref{eq:FIM_coop} becomes a block-diagonal matrix of blocks $\mathbf{N}_m$ and the inverse FIM a block-diagonal matrix of blocks $\mathbf{N}_m^{-1}$. Thus, in the non-cooperative case the PEB simplifies to
\begin{align}
\PEB_m \ncoop = \sqrt{ \sum_{k=1}^3 \ \left [ \,\mathbf{N}_m \inv \,\right ] _{k,k} } \eqend
\end{align}

%% file: 5_Evaluation.tex
\section{Numerical Evaluation}
\label{sec:eval}
The following numerical evaluation considers the cubic setup depicted in \Cref{fig:3DRoom}. All \TA subcoils are chosen to be identical with $\nu=5$ wire turns and a diameter of $D=\SI{5}{\centi \meter}$, which leads to a surface area of $A=\SI{6.25\pi}{\centi \meter \squared}$. We use an operating frequency of $f=\SI{500}{\kilo \hertz}$ and assume that $\mu = 4\pi \times 10^{-7} \; \mathrm{H}\mathrm{m}^{-1}$. We assume a measurement error standard deviation of $\sigma=10^{-5}$. This  is in accordance with \cite{bib:dumphart2019practical} who found this error level to be appropriate for office environments, based on vector network analyzer measurements with coils similar to ours. For practical applications which perform the channel estimation via wireless pilot signals, the measurement error variance may be higher, leading to an overall decreased accuracy. For all required numerical optimization steps we use the Levenberg-Marquadt algorithm \cite{bib:levenberg1944method} via  Matlab's built-in \textit{lsqnonlin} function \cite{bib:lsqnonlin}.  We consider $\N=4$ anchors with the same fixed positions at the boundaries for all setups (cf. \Cref{fig:3DRoom}) and a variable number $\M$ of agents. The agents' placement and orientation is uniformly random within the room but restricted such that it satisfies a minimum distance constraint ($r\csub \geq 3D$) between any coil pair. Lastly, all used performance measures are taken with respect to agent $1$. However, the performance of any other agent is statistically identical to that of agent $1$,  so this specific choice does not affect the generality.

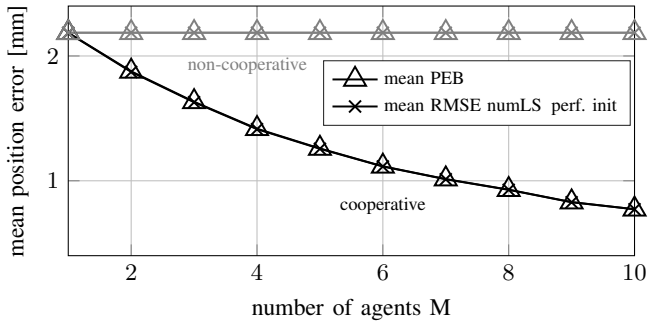
\begin{figure}
\centering
\input{CoopGain.tex}
\caption{Comparison of the localization accuracy for cooperative vs. \NC approaches with an increasing number of agents in the room.  The numerical ML estimators are perfectly initialized and their resulting RMSEs tightly approach the corresponding PEBs.}
\label{fig:RMSEvsNag}
\end{figure}

\begin{figure}
\centering
\input{CDF_channelCoeff.tex}
\caption{CDFs of the absolute values of inter-node channel coefficients for $\M=10$ randomly deployed agents in the setup.}
\label{fig:CDF_channelCoeff}
\end{figure}
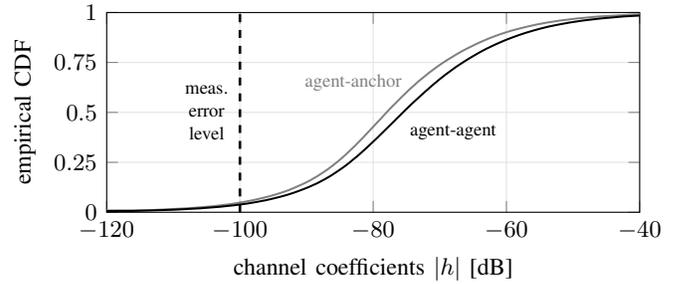

\Cref{fig:RMSEvsNag} shows how the accuracy for both cooperative (black) and \NC (gray) localization schemes changes with an increasing number of agents.  For each number of agents $\M$, $1000$ random agent topologies, each with $100$ different measurement error realizations, were considered. The triangle-lines represent the corresponding mean PEBs of agent $1$ for each $\M$, whereas the cross-lines illustrate the mean RMSEs of the corresponding LS estimators, averaging over agent topologies and measurement error realizations. The numerical LS estimators (\numLS) used the true agent deployment as starting points.  We see that the minimum which is found with this perfect initialization corresponds to the PEB. Without cooperation between the agents, unsurprisingly, the performance is not affected by their number $\M$ and remains constant. For the cooperative scenario we see a steady improvement with increasing $M$. This can be attributed both to having more channel observations and to having stronger channel gains between agent pairs that are in close proximity. In \Cref{fig:CDF_channelCoeff} this aspect is further emphasized by the cumulative distribution functions (CDFs) of the channel gains $\left | h \right |$ between any pair of either agent-agent or agent-anchor subcoils for $\M=10$ randomly placed agents within the room. We observe that about $4 \%$ of the channel gains are below the measurement error level and that the agent-agent channel gains are generally higher than the channel gains between agents and anchors. 

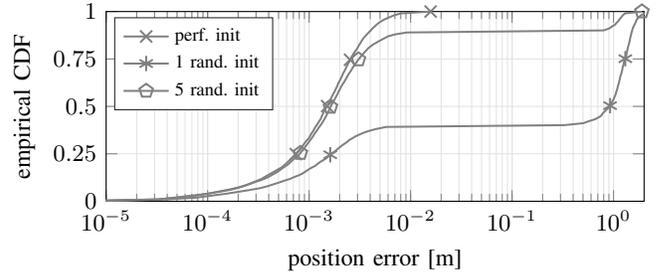
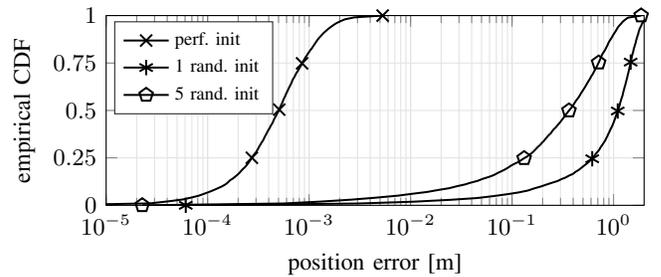
\begin{figure}
\begin{subfigure}[b]{0.95\columnwidth}
\centering
\input{CDF_NCoop_RI.tex}
\caption{Non-cooperative least-squares estimation}
\label{fig:CDFs_subfig_ncoop}
\end{subfigure}
\begin{subfigure}[b]{0.95\columnwidth}
\centering
\input{CDF_Coop_RI.tex}
\caption{Cooperative least-squares estimation}
\label{fig:CDFs_subfig_coop}
\end{subfigure}
\caption{CDFs of the instantaneous localization errors achieved by the numerical least-squares estimation (\numLS) for different initializations.}
\label{fig:CDFs_inst_error}
\end{figure}

Thanks to the initialization with the true deployment vector, the numerical solvers in \Cref{fig:RMSEvsNag} approach the theoretical limit constituted by the PEB. Without such unrealistic perfect initialization however, it remains unclear whether the cooperation gain is actually feasible. To investigate this matter,  \Cref{fig:CDFs_inst_error} shows the corresponding CDFs of the Euclidean norms of all instantaneous positions errors for the case of $\M=10$ randomly deployed agent nodes (using many different realizations of the agent topology and the measurement error). 
In \Cref{fig:CDFs_subfig_ncoop,fig:CDFs_subfig_coop}, we distinguish between the perfect initialization (\textit{perf. init.}) and a random initialization (\textit{rand. init.}) within the room with either $1$ or $5$ random initializations\footnote{Ultimately we chose the estimate with the lowest residual cost (i.e. the lowest squared error between measurements and model) over all individual Levenberg-Marquardt solver runs from the different initial values.}. For the easier \NC localization in \Cref{fig:CDFs_subfig_ncoop}, we observe that a single random initialization only suffices in about $40 \%$ of the cases and otherwise converges to local minima. With $5$ random initializations, only $10 \%$ of these outliers remain and with even more random initializations the outliers could possibly be rendered insignificant. For the cooperative localization in \Cref{fig:CDFs_subfig_coop}, we see that additional random initializations also generally improve the performance, but even with $5$ random initializations we still do not even remotely approach the perfectly initialized case. Even worse, despite combining cooperation and multiple random initializations, the performance is worse than the one of the corresponding \NC approach. Moreover, using five initializations entails a five-fold complexity increase, which is especially problematic for the high-dimensional cooperative joint optimization. This shows that the increased computational complexity can effectively nullify the cooperation gain which the additional agent-agent observations otherwise provide. 

In \Cref{fig:CDFs_subfig_3step}, we thus additionally compare the performance of the proposed estimators from \Cref{sec:localization}, which are briefly summarized in \Cref{tab:localization_summary}.
\begin{table}[tb]
\vspace{0.04in}
\caption{Summary of all proposed localization schemes. }
\label{tab:localization_summary}
\begin{tabular}{lp{6.3cm}}
Scheme &\multicolumn{1}{l}{Description} \\ \hline
\\[-1.5mm]
%
\numLS & A numerical solver (Levenberg-Marquadt algorithm) with a certain initial parameter value is applied to the cost function from either \Cref{eq:localization_LS_coop} or \Cref{eq:localization_LS_noncoop} in an attempt to solve the associated LS problem.\\
\\[-1.5mm] 
\pairML & Position and orientation estimates of an agent are calculated via the closed-form ML estimation rules  \Cref{eq:pairwiseMLorientation}, \Cref{eq:pairwiseMLposition}, based on the measurement of the anchor with the smallest ML-estimated distance (cf. \Cref{subsec:advancedInit}).\\
\\[-1.5mm]
\turboLS & The position and orientation estimates obtained with \pairML \, are used as initial parameter values in \numLS, which functions as afterburner. \\
\\[-1.5mm]
multilateration  & Obtain ML distance estimates between an agent and all anchors with the closed-form rule \Cref{eq:pairwiseMLdistance}. Then estimate the agent position via multilateration (a gradient search).
%
\end{tabular}
\end{table}
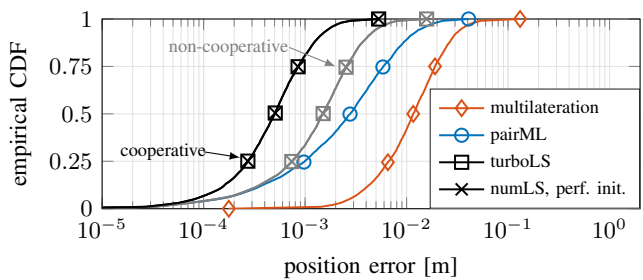
\begin{figure}
\begin{subfigure}[b]{0.95\columnwidth}
\centering
\input{CDF_3Step_Multilat.tex}
\end{subfigure}
\caption{CDF comparison of the instantaneous localization errors achieved for all proposed localization estimators.}
\label{fig:CDFs_subfig_3step}
\end{figure}
For both the cooperative and the \NC scenarios, the \turboLS \, scheme (squares) achieves the same performance as the perfectly initialized \numLS \, scheme (crosses). Without the LS afterburner, the \pairML \, scheme (circles) maintains an acceptable performance that is comparable to that of the \NC scheme despite its ultra-low complexity and despite only requiring a single agent-anchor link.
Lastly, we consider the common approach of position estimation via multilateration, which processes distance estimates. Specifically, we use all available agent-anchor distance estimates $\hat{r}\csub^\mathrm{ML}$ from \eqref{eq:pairwiseMLdistance} to determine the position hypothesis $\hat{\mathbf{p}}_\m$ which minimizes
$\sum_{n=\M + 1}^\N (r\csub(\mathbf{p}_\m) - \hat{r}\csub^\mathrm{ML})^2$ with a gradient search. We find that multilateration (diamonds) exhibits poor accuracy compared to all other approaches.

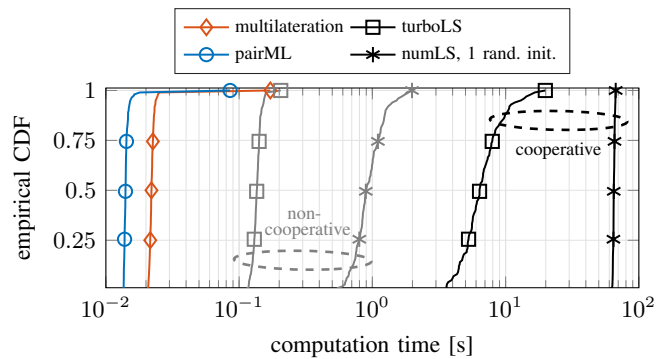
\begin{figure}
\centering
\input{CDF_time.tex}
\caption{CDF comparison of the time which each approach requires to fully localize a network comprising $\M=10$ agents.}
\label{fig:CDF_time}
\end{figure}

In order to further emphasize on the computational differences, \Cref{fig:CDF_time} shows the CDFs of the time that each approach requires in order to fully localize $\M=10$ agents. To this end, the optimization processes were run on an Intel Core i7-10750H  $\SI{2.6}{\giga \hertz}$ processor in the single core mode. 
For this processor, the \pairML \, approach requires less than $\SI{18}{\milli \second}$ in almost all cases and the multilateration less than $\SI{25}{\milli \second}$. For the LS approaches, we observe that the \NC \turboLS \, has a median processing time of $\SI{136}{\milli \second}$ and its cooperative equivalent requires roughly $\SI{6}{ \second}$. We also see that these \turboLS \, approaches, which are required to reliably find the global minimum, reduce the median processing time compared to the \numLS \, solution with a single random initialization. Nevertheless, the required processing time of any cooperative approach compared to its \NC alternative is still increased by more than an order of magnitude. Hence, cooperative localization may not be viable for time-critical applications with contemporary computing hardware. In combination with \Cref{fig:CDFs_subfig_3step} we observe a clear trade-off between the improved localization accuracy and the increased computational complexity that additional cooperation entails. 

%% file: CoopGain.tex
\definecolor{mycolor1}{rgb}{0.85098,0.32549,0.09804}%
\definecolor{mycolor2}{rgb}{0.00000,0.44706,0.74118}%
\begin{tikzpicture}

\begin{axis}[%
width=0.85\columnwidth,
height=0.375\columnwidth,,
scale only axis,
xmin=1,
xmax=10,
xlabel={number of agents M},
ymin=0.4,
ymax=2.4,
ylabel={mean position error [mm]},
axis background/.style={fill=white},
xmajorgrids,
ymajorgrids,
legend style={at={(0.45,0.52)}, anchor=south west, legend cell align=left, align=left, draw=white!15!black, font =\scriptsize}
]
\plotPEB{black}
table[row sep=crcr]{%
16	1.80722090365559\\
};
\addlegendentry{mean PEB}

\plotPinit{black}{}
table[row sep=crcr]{%
16	1.80058313581985\\
};
\addlegendentry{mean RMSE \numLS \, perf. init}

\plotPEB{black}
table[row sep=crcr]{%
1	2.18627459283404\\
2	1.87682145847361\\
3	1.63302904392104\\
4	1.41731290160644\\
5	1.25907761606583\\
6	1.11624135224503\\
7	1.01375733242998\\
8	0.928937128504422\\
9	0.829387886520207\\
10	0.772151148042094\\
};

\plotPinit{black}{forget plot}
table[row sep=crcr]{%
1	2.1839318710071\\
2	1.8721567062113\\
3	1.62928793274131\\
4	1.41350088605878\\
5	1.2562717354954\\
6	1.11462162281033\\
7	1.01224979874621\\
8	0.927490165247549\\
9	0.828082875540121\\
10	0.772033049429613\\
};

\plotPEB{matGray}
table[row sep=crcr]{%
1	2.18627459283404\\
2	2.18627459283404\\
3	2.18627459283404\\
4	2.18627459283404\\
5	2.18627459283404\\
6	2.18627459283404\\
7	2.18627459283404\\
8	2.18627459283404\\
9	2.18627459283404\\
10	2.18627459283404\\
};

\plotPinit{matGray}{forget plot}
table[row sep=crcr]{%
1	2.18627459283404\\
2	2.18627459283404\\
3	2.18627459283404\\
4	2.18627459283404\\
5	2.18627459283404\\
6	2.18627459283404\\
7	2.18627459283404\\
8	2.18627459283404\\
9	2.18627459283404\\
10	2.18627459283404\\
};

\node[align=center, font=\color{matGray}] at (3.9,1.92) {\scriptsize \NC};
\node[align=center, font=\color{black}] at (6,0.8) {\scriptsize cooperative};
\end{axis}
\end{tikzpicture}%

%% file: CDF_channelCoeff.tex
%
%
\begin{tikzpicture}

\begin{axis}[%
width=0.8\columnwidth,
height=0.3 \columnwidth,
scale only axis,
xmin=-120,
xmax=-40,
xlabel={channel coefficients $|h|$ [dB]},
ymin=0,
ymax=1,
ytick={  0,  0.25, 0.5, 0.75, 1},
ylabel={empirical CDF},
axis background/.style={fill=white},
xmajorgrids,
ymajorgrids,
grid style={gridGray}
]

\addplot [color=matGray, line width=\mylinewidth]
  table[row sep=crcr]{%
-177.276769521141	0\\
-113.600761066863	0.0101296296296306\\
-107.595409792408	0.0203009259259262\\
-103.999644951217	0.0302175925925935\\
-101.531635776105	0.0401481481481509\\
-99.6029347806676	0.0503796296296365\\
-98.0388673821698	0.0605324074074192\\
-96.6972834644669	0.0706898148148261\\
-95.5134494124421	0.0806435185185302\\
-94.4918146560877	0.0905416666666811\\
-93.541222975253	0.100666666666682\\
-92.6991574108359	0.11065740740743\\
-91.968919553296	0.120699074074101\\
-91.271091908802	0.130750000000032\\
-90.6002158254347	0.140981481481516\\
-89.9640401928161	0.151032407407448\\
-89.3751691899917	0.161171296296346\\
-88.8579044338756	0.171324074074129\\
-88.3377310159867	0.181486111111172\\
-87.868628919451	0.191722222222291\\
-87.401105556444	0.20182870370378\\
-86.9522771926569	0.21206944444453\\
-86.5425082781667	0.222115740740834\\
-86.1356196504566	0.232189814814916\\
-85.752260611802	0.242421296296407\\
-85.3766023222175	0.252523148148269\\
-85.0083246605615	0.262500000000133\\
-84.6654433517573	0.272509259259403\\
-84.3303628951615	0.282754629629787\\
-83.9960582958419	0.292675925926092\\
-83.6754368214649	0.302837962963145\\
-83.3623313177181	0.312814814815009\\
-83.0432574351613	0.323069444444654\\
-82.7150603878023	0.333004629629851\\
-82.4258151224579	0.343310185185422\\
-82.1273577301796	0.353268518518772\\
-81.8286485035026	0.363337962963234\\
-81.5322924508574	0.373240740741024\\
-81.2595873168647	0.383296296296597\\
-80.9802249953904	0.393388888889204\\
-80.6902748321619	0.403560185185515\\
-80.4025092590202	0.413833333333679\\
-80.1105415761389	0.423879629629992\\
-79.8264961161063	0.434092592592972\\
-79.5458870255134	0.444208333333734\\
-79.2555270212224	0.454416666667087\\
-78.9696442916744	0.464444444444881\\
-78.6897701179687	0.474527777778237\\
-78.4122813748578	0.484800925926402\\
-78.1304337462643	0.494912037037532\\
-77.847550179635	0.505050925926426\\
-77.5534167054502	0.515277777778267\\
-77.2642796897568	0.525337962963439\\
-76.964926226997	0.535587962963427\\
-76.6541757648499	0.545643518518971\\
-76.3591368686305	0.555587962963404\\
-76.0572503445371	0.565708333333762\\
-75.759036022983	0.575879629630047\\
-75.4425144097977	0.586037037037443\\
-75.1296799614268	0.595953703704097\\
-74.8171417188578	0.606310185185566\\
-74.5013304771119	0.616388888889258\\
-74.1643586867978	0.62657870370406\\
-73.8315036285734	0.636490740741084\\
-73.4923297840648	0.646662037037367\\
-73.1529210319474	0.65683796296328\\
-72.7787931959287	0.666879629629933\\
-72.4025307723412	0.677004629629919\\
-72.0292093152932	0.687087962963237\\
-71.6584208097702	0.697240740741001\\
-71.2664742214554	0.707254629629877\\
-70.8580036110038	0.717328703703937\\
-70.443248540767	0.727513888889109\\
-70.0118231637754	0.737587962963169\\
-69.570261551841	0.747787037037228\\
-69.1125348911344	0.757745370370552\\
-68.6279535929197	0.767717592592769\\
-68.1304575961479	0.777699074074244\\
-67.6401943095875	0.787791666666831\\
-67.1266635156687	0.79785185185201\\
-66.5814946229478	0.808023148148301\\
-65.996844885423	0.818226851852001\\
-65.3899701226075	0.828305555555701\\
-64.7726854796044	0.838393518518661\\
-64.1310620523004	0.848671296296436\\
-63.425815148178	0.858953703703841\\
-62.7203608536163	0.86918518518532\\
-61.9290708110206	0.879305555555686\\
-61.114979903255	0.889347222222342\\
-60.2060617782731	0.899300925926035\\
-59.2393637069082	0.909513888888987\\
-58.0828573636473	0.919541666666754\\
-56.8420818934139	0.92969907407415\\
-55.4755744614272	0.939953703703768\\
-53.8657754343241	0.950013888888943\\
-51.9086116564598	0.960101851851896\\
-49.4524241562747	0.970120370370405\\
-46.1284241022314	0.980240740740764\\
-40.8130596336868	0.990120370370382\\
-19.9218837348279	1\\
};

\addplot [color=black, line width=\mylinewidth]
  table[row sep=crcr]{%
-192.330918668336	0\\
-116.116988917305	0.00625102880658535\\
-109.831259254151	0.0128106995884821\\
-105.938853953534	0.0197325102880692\\
-103.10734844737	0.0271399176954789\\
-100.908394115946	0.0348477366255258\\
-99.005127515332	0.0429917695473389\\
-97.4828622299177	0.0516008230452853\\
-96.1129492321651	0.0602386831275958\\
-94.9532071267427	0.0691522633745139\\
-93.8718554131104	0.0782263374485933\\
-92.8774968881895	0.0875185185185615\\
-91.99702455949	0.096913580246961\\
-91.1984523602079	0.106403292181126\\
-90.4385546892669	0.115979423868375\\
-89.7105594333455	0.125707818930111\\
-89.0573929283984	0.135386831275798\\
-88.4042741804119	0.145296296296383\\
-87.8360088268815	0.155213991769643\\
-87.2756055900502	0.165135802469243\\
-86.7501522884924	0.175234567901354\\
-86.2356560100101	0.185551440329347\\
-85.7768503559216	0.195683127572155\\
-85.3228250617241	0.205761316872581\\
-84.8757541255705	0.215983539094819\\
-84.4326318370308	0.226127572016646\\
-84.0244974267225	0.236582304526949\\
-83.6176542404984	0.247053497942604\\
-83.2330177898389	0.25770370370394\\
-82.8515287689613	0.268197530864449\\
-82.4626453455819	0.278604938271875\\
-82.100965926728	0.289028806584652\\
-81.7485977370097	0.299555555555866\\
-81.4011172352543	0.310039094650535\\
-81.0441535134405	0.320798353909812\\
-80.6989120823368	0.331526748971559\\
-80.3660053894426	0.342037037037423\\
-80.0218005257851	0.352777777778186\\
-79.6944414910222	0.363234567901669\\
-79.361562148438	0.373528806584814\\
-79.052710764463	0.383925925926403\\
-78.7308924635952	0.39438683127622\\
-78.4156863206226	0.404699588477891\\
-78.1022567966879	0.415000000000546\\
-77.7819837739666	0.425526748971766\\
-77.4663741928862	0.435967078189897\\
-77.1455896379815	0.446506172840126\\
-76.8191294980078	0.457185185185828\\
-76.5070202317007	0.467802469136472\\
-76.1862826939074	0.478325102881353\\
-75.8596613936251	0.48900000000072\\
-75.539509220699	0.499613168725027\\
-75.2238629820139	0.510131687243532\\
-74.8943173434857	0.520720164609773\\
-74.5691043148684	0.531069958848439\\
-74.237437907194	0.541695473251716\\
-73.8932690243578	0.552255144033593\\
-73.5495615177596	0.562909465021231\\
-73.2042884055025	0.573358024691995\\
-72.8617339993851	0.583905349794858\\
-72.5034818029375	0.594423868313359\\
-72.1576821097017	0.605172839506759\\
-71.7876890965114	0.61574279835448\\
-71.4356869771437	0.626345679012899\\
-71.0643529139213	0.637069958848273\\
-70.6818397766939	0.647753086420274\\
-70.2949443894588	0.658526748971697\\
-69.9071073835517	0.669209876543697\\
-69.4994229126216	0.679827160494296\\
-69.0922306043669	0.690510288066295\\
-68.6443785186352	0.701308641975745\\
-68.2057518208474	0.71202880658478\\
-67.7688625433081	0.722777777778176\\
-67.3039101261621	0.733502057613547\\
-66.822453643747	0.743942386831637\\
-66.3325037791803	0.754534979424214\\
-65.8272322327622	0.765226337448891\\
-65.2990807880541	0.775843621399494\\
-64.7594730565188	0.786456790123761\\
-64.1934654010093	0.797259259259551\\
-63.6420063732384	0.807572016461185\\
-63.0672851628739	0.817588477366525\\
-62.4525785317783	0.82758024691384\\
-61.8115350076363	0.837691358024942\\
-61.1334076463352	0.847880658436457\\
-60.4325864584191	0.857962962963197\\
-59.6725590062974	0.868213991769773\\
-58.8527190465392	0.87842386831297\\
-58.0143920140687	0.888438271605134\\
-57.0833046055186	0.89865843621417\\
-56.0228065938839	0.909127572016622\\
-54.9186206629271	0.919358024691502\\
-53.7249803466296	0.929279835391074\\
-52.2661960803441	0.939246913580359\\
-50.6014309428649	0.949436213991862\\
-48.5390027652484	0.959847736625588\\
-45.9437610685091	0.970016460905404\\
-42.4581542570136	0.980275720164644\\
-37.0162562815066	0.990411522633762\\
-18.429852541199	1\\
};

\addplot [color=black, dashed, line width=1.0pt]
  table[row sep=crcr]{%
-100	0\\
-100	1\\
};
\node[align=center, font=\color{matGray}] at (-83,0.65) {\scriptsize agent-anchor};
\node[align=left, font=\color{black}] at (-68,0.4) {\scriptsize agent-agent};
\node[align=center, font=\color{black}] at (-105,0.5) {\scriptsize meas.\\[-1.2mm]\scriptsize error\\[-1.2mm]\scriptsize level};
\end{axis}

\end{tikzpicture}%

%% file: CDF_NCoop_RI.tex
%
%
\definecolor{mycolor1}{rgb}{0.00000,0.44706,0.74118}%
\begin{tikzpicture}

\begin{axis}[%
width=0.85\columnwidth,
height=0.3 \columnwidth,
scale only axis,
xmode=log,
xmin=1e-05,
xmax=2,
xminorticks=true,
xlabel={position error [m]},
ymin=0,
ymax=1,
ytick={  0,  0.25, 0.5, 0.75, 1},
ylabel={empirical CDF},
axis background/.style={fill=white},
xmajorgrids,
xminorgrids,
ymajorgrids,
grid style={gridGray},
legend style={at={(0.015,0.97)}, anchor=north west, legend cell align=left, align=left, draw=white!15!black, font=\scriptsize}
]


\plotPinit{matGray}{}
table[row sep=crcr]{%
100	1\\
};
\addlegendentry{perf. init}

\plotPinit{matGray}{forget plot, mark=none}
table[row sep=crcr]{%
3.2842863438936e-06	0\\
3.27734044777462e-05	0.0101666666666663\\
4.99809183392531e-05	0.0208333333333327\\
7.15215178632704e-05	0.0303333333333332\\
9.86656212658874e-05	0.0398333333333333\\
0.000128149246980109	0.0496666666666664\\
0.000165011138421207	0.0596666666666662\\
0.00019693745608415	0.0695\\
0.000223430479058421	0.0793333333333337\\
0.000250367455665966	0.0890000000000007\\
0.000277898643862753	0.0990000000000008\\
0.000309419670925857	0.108833333333333\\
0.000341014412939451	0.118333333333333\\
0.000365749342727433	0.128166666666666\\
0.000393598466421361	0.137833333333333\\
0.000423718850766218	0.147666666666666\\
0.000457283564647887	0.157166666666666\\
0.000491281003737809	0.167333333333332\\
0.000531225254289406	0.176999999999998\\
0.00056074065829462	0.186999999999997\\
0.000591509784587231	0.197666666666663\\
0.000619076509069033	0.208166666666663\\
0.000648210475086847	0.217999999999995\\
0.000677404167757512	0.227999999999995\\
0.000710808342628737	0.237999999999994\\
0.000740195338155888	0.24866666666666\\
0.000764978705469697	0.258499999999993\\
0.000796436022380328	0.26866666666666\\
0.000820540970640621	0.278833333333326\\
0.000854840694668714	0.289166666666659\\
0.000885750455150127	0.300166666666658\\
0.000917145025008286	0.311166666666657\\
0.000950355509758901	0.321333333333323\\
0.000990943240354239	0.331333333333322\\
0.00101543483617363	0.341499999999988\\
0.00104039546744355	0.351666666666655\\
0.00107012719686915	0.361999999999987\\
0.00110299920102573	0.372166666666653\\
0.00112773689687534	0.382166666666652\\
0.00115520884473625	0.392166666666651\\
0.00119145127718432	0.401833333333317\\
0.00122683429599596	0.411833333333316\\
0.00125497124891152	0.421499999999982\\
0.00129377211849664	0.431499999999981\\
0.00131790888436846	0.441333333333313\\
0.00134481597653174	0.451999999999979\\
0.00137276317086137	0.462499999999978\\
0.00140454007366901	0.47233333333331\\
0.00143361302426257	0.482166666666643\\
0.00146348329999766	0.491999999999975\\
0.00150697414514773	0.501666666666641\\
0.0015378031619711	0.511833333333308\\
0.00157257921141179	0.521999999999975\\
0.00161113480277512	0.531666666666642\\
0.00164199368005524	0.541333333333309\\
0.00167355512004456	0.551166666666643\\
0.00171230486713249	0.560999999999977\\
0.00175397417523507	0.571166666666644\\
0.00179442389960523	0.581166666666644\\
0.00183050483354576	0.591499999999978\\
0.00187286709235398	0.601499999999979\\
0.00191066468761937	0.611999999999979\\
0.00193655085951919	0.622166666666646\\
0.00198785587277208	0.631666666666647\\
0.00203180300313122	0.642166666666648\\
0.00207588508230415	0.651999999999981\\
0.00211879380318111	0.662666666666648\\
0.00215905872419039	0.672833333333315\\
0.00221430257652085	0.683333333333315\\
0.00227120945979942	0.69416666666665\\
0.00231882198961029	0.704833333333317\\
0.00236669445610214	0.71466666666665\\
0.00241804807544971	0.724999999999985\\
0.00248585835647708	0.734999999999985\\
0.00253343399557336	0.745833333333319\\
0.00259069293028403	0.755999999999986\\
0.00265772154405582	0.76583333333332\\
0.0027288846679025	0.776166666666654\\
0.0028095021561273	0.786333333333321\\
0.00290515323427043	0.796833333333322\\
0.00299640068689915	0.806999999999989\\
0.00305963077036763	0.817166666666656\\
0.00314073547273102	0.827166666666657\\
0.00323417345337976	0.837666666666657\\
0.00331817576709091	0.847833333333325\\
0.00342314601348122	0.858499999999992\\
0.00355977748301921	0.869166666666659\\
0.0037056658994467	0.879333333333326\\
0.00386942108878251	0.88966666666666\\
0.00404616477689709	0.899833333333328\\
0.00419000963475914	0.909833333333328\\
0.00439804985629735	0.919499999999995\\
0.00459620800139127	0.929499999999996\\
0.00483814149012222	0.939499999999997\\
0.0051706567460741	0.949666666666664\\
0.00556694732504295	0.959333333333331\\
0.00620248978255262	0.969833333333332\\
0.00684250352659337	0.979833333333332\\
0.00803425670527542	0.990333333333333\\
0.0157524026609455	1\\
};

\plotPinit{matGray}{forget plot, only marks}
table[row sep=crcr]{%
3.2842863438936e-06	0\\
0.000740195338155888	0.24866666666666\\
0.00150697414514773	0.501666666666641\\
0.00253343399557336	0.745833333333319\\
0.0157524026609455	1\\
};


\plotRinitOne{matGray}{}
  table[row sep=crcr]{%
100	1\\
};
\addlegendentry{1 rand. init}

\plotRinitOne{matGray}{forget plot, no marks}
  table[row sep=crcr]{%
6.18489327941637e-06	0\\
3.86464081685258e-05	0.0093333333333333\\
7.78605150492611e-05	0.0193333333333328\\
0.000107261003735891	0.029166666666666\\
0.000154649482393271	0.0401666666666659\\
0.000203394347475068	0.0503333333333326\\
0.000264087767872083	0.0599999999999989\\
0.000331839010169006	0.0694999999999991\\
0.000389740117279621	0.0793333333333325\\
0.000444761009122575	0.0893333333333328\\
0.000509017431726724	0.0991666666666662\\
0.00056975083484634	0.108499999999999\\
0.000647857293290164	0.117666666666666\\
0.000713453640937197	0.127666666666665\\
0.000797613761632599	0.137499999999998\\
0.000865132689936809	0.146999999999998\\
0.000925724599285845	0.157166666666665\\
0.000983537055321418	0.166833333333331\\
0.00105461567978271	0.176499999999997\\
0.00113956630855271	0.185666666666664\\
0.00120735635676998	0.195999999999996\\
0.0012826889908823	0.206833333333329\\
0.00137522645015803	0.216333333333329\\
0.00146859037242224	0.225333333333328\\
0.00154999525617589	0.235833333333328\\
0.00161600819586037	0.24516666666666\\
0.00169850411626178	0.254833333333326\\
0.00181359672170811	0.264999999999993\\
0.00189841744125059	0.274333333333325\\
0.00199321424503428	0.283666666666658\\
0.00210807577391794	0.293999999999991\\
0.00223324686855526	0.303166666666657\\
0.00236878452930394	0.312666666666656\\
0.00247290658803863	0.321833333333322\\
0.0026468493104757	0.330999999999989\\
0.00284684579726147	0.341333333333321\\
0.0031028711190456	0.350666666666654\\
0.00344232928713808	0.360999999999987\\
0.00389615394010098	0.37083333333332\\
0.00455935891879562	0.380666666666652\\
0.00575546781521897	0.391833333333318\\
0.324679981470517	0.40116666666665\\
0.501697851197947	0.411333333333316\\
0.577940503924607	0.422999999999982\\
0.65771526574325	0.432166666666648\\
0.713234608090561	0.442166666666647\\
0.757113146331094	0.451166666666646\\
0.779960375425604	0.462999999999979\\
0.821164373765618	0.474166666666645\\
0.854594309359927	0.483833333333311\\
0.898404123046687	0.494499999999977\\
0.924863145747884	0.503999999999977\\
0.945427242057739	0.514166666666644\\
0.969612553163819	0.523999999999978\\
0.996439487528779	0.535166666666645\\
1.01307016775425	0.544499999999979\\
1.03263616490193	0.555166666666646\\
1.04973463486947	0.564166666666646\\
1.0610886059311	0.57399999999998\\
1.076723278993	0.584999999999981\\
1.09265390393849	0.595999999999981\\
1.10279916004519	0.606166666666648\\
1.11710353746057	0.615833333333315\\
1.13168009538465	0.625499999999982\\
1.14090820789631	0.634833333333316\\
1.15175797087393	0.64366666666665\\
1.16324155778329	0.65416666666665\\
1.17475477927829	0.662999999999984\\
1.19184842635894	0.671833333333318\\
1.2033141441074	0.682166666666651\\
1.21869935798733	0.691833333333318\\
1.22600890539675	0.701999999999985\\
1.2381393424135	0.711666666666652\\
1.25654127272645	0.721166666666653\\
1.2650710995667	0.73083333333332\\
1.27990141930705	0.74083333333332\\
1.30602475945702	0.751499999999987\\
1.3239485212601	0.761333333333321\\
1.33776591381331	0.770499999999988\\
1.34575962988719	0.781666666666656\\
1.36106577384124	0.791333333333323\\
1.37586311819432	0.802833333333323\\
1.38762662486047	0.812166666666657\\
1.40796524829854	0.823666666666658\\
1.42573068842374	0.833833333333325\\
1.43978455197879	0.843999999999992\\
1.45505716816841	0.855999999999993\\
1.46920298249591	0.866999999999994\\
1.4884104059095	0.878499999999994\\
1.50434999220632	0.890333333333328\\
1.52674768631148	0.899666666666662\\
1.54167345346381	0.909999999999996\\
1.5826311796838	0.923999999999997\\
1.62593026865997	0.935666666666664\\
1.65720448619156	0.946166666666664\\
1.69216288965555	0.957333333333332\\
1.75887441168001	0.968499999999999\\
1.85450847381362	0.978999999999999\\
1.98969670342116	0.988666666666666\\
2.37588871393175	1\\
};

\plotRinitOne{matGray}{forget plot, only marks}
  table[row sep=crcr]{%
6.18489327941637e-06	0\\
0.00161600819586037	0.24516666666666\\
0.924863145747884	0.503999999999977\\
1.30602475945702	0.751499999999987\\
2.37588871393175	1\\
};


\plotRinitFive{matGray}{}
  table[row sep=crcr]{%
100 1\\
};
\addlegendentry{5 rand. init}

\plotRinitFive{matGray}{forget plot, no marks}
  table[row sep=crcr]{%
6.18488775898091e-06	0\\
3.22308792681036e-05	0.0096666666666666\\
5.27925522629264e-05	0.0201666666666666\\
7.8104431616132e-05	0.0306666666666668\\
0.000104290875255349	0.0411666666666669\\
0.000132174525097968	0.0523333333333335\\
0.000167416647762499	0.0628333333333334\\
0.000197815536765101	0.0721666666666669\\
0.000233215313513735	0.0825000000000006\\
0.000266620953860383	0.0930000000000004\\
0.000300454977156104	0.103166666666667\\
0.000340824596517846	0.112666666666666\\
0.000380571302213885	0.122999999999999\\
0.000412796756627185	0.133166666666666\\
0.000444514166540939	0.142999999999999\\
0.000482468385151732	0.152999999999999\\
0.000512241954131489	0.163166666666665\\
0.000543260734644352	0.173499999999998\\
0.000579595966619763	0.183333333333331\\
0.000618158547602406	0.193666666666664\\
0.000647888630752186	0.203666666666663\\
0.000677443176786066	0.213499999999996\\
0.000713977434220862	0.224166666666662\\
0.000758506184194265	0.233999999999995\\
0.000790535648026026	0.243499999999994\\
0.000821022654975755	0.254166666666661\\
0.000849742919079294	0.26416666666666\\
0.000883784888005998	0.274499999999993\\
0.000906462151952476	0.284333333333326\\
0.000937599053018944	0.294499999999992\\
0.000970156693332374	0.304666666666658\\
0.00100220402693215	0.314333333333324\\
0.00103351934230563	0.324333333333324\\
0.0010682016906857	0.334833333333323\\
0.00110046804369498	0.344999999999989\\
0.00112972804855403	0.355499999999989\\
0.00116625028912254	0.365833333333321\\
0.00120520398769848	0.375833333333321\\
0.00123641008138532	0.385999999999987\\
0.00128085497017604	0.396666666666652\\
0.00132206438974693	0.406999999999985\\
0.00135861299469096	0.416833333333317\\
0.00139361792465145	0.426833333333316\\
0.00142824406758447	0.436999999999982\\
0.00146182649094021	0.446666666666648\\
0.00149209455178022	0.45599999999998\\
0.00152422524409877	0.466333333333313\\
0.0015626091365794	0.476499999999978\\
0.00159462103348944	0.486833333333311\\
0.00162332794697027	0.49733333333331\\
0.00166159497845995	0.507166666666644\\
0.00169751777319077	0.516999999999977\\
0.00174687562617435	0.526999999999978\\
0.00178553126968099	0.537333333333312\\
0.00182724147108766	0.547333333333312\\
0.00187239694050753	0.557833333333313\\
0.00191721427778006	0.568333333333313\\
0.00195407294420336	0.57899999999998\\
0.00200548920793165	0.588999999999981\\
0.00204845074121429	0.599333333333315\\
0.00210046071368813	0.609166666666648\\
0.00215001894433387	0.619166666666649\\
0.00221665542614829	0.628499999999983\\
0.00226815681235119	0.638499999999984\\
0.00232530035136051	0.64816666666665\\
0.00237739190404614	0.657499999999984\\
0.00243340769957776	0.667166666666651\\
0.00249856492627686	0.676999999999985\\
0.0025834517275125	0.687499999999985\\
0.00264243085092077	0.696666666666652\\
0.00270360351820619	0.706666666666652\\
0.0027946089394746	0.716833333333319\\
0.0028775696930135	0.72733333333332\\
0.00297292213170222	0.737666666666654\\
0.00306616224214895	0.747166666666654\\
0.00316419719739713	0.756499999999988\\
0.00328335845437257	0.767166666666655\\
0.00340666507518366	0.777333333333322\\
0.00355351603505133	0.787666666666656\\
0.00370500689406587	0.79799999999999\\
0.00384786170488377	0.807833333333324\\
0.00405802050163004	0.817166666666658\\
0.00429642648073295	0.827166666666658\\
0.00466213546183566	0.837666666666659\\
0.0050386132508364	0.847999999999993\\
0.00548596556302489	0.858333333333327\\
0.0060603925204028	0.870333333333327\\
0.00709100171989853	0.879833333333328\\
0.00921133343682099	0.889666666666662\\
0.796094408940474	0.900499999999996\\
0.902795513185359	0.910499999999996\\
0.963430816172375	0.92083333333333\\
1.03239319871556	0.930666666666664\\
1.07197023776074	0.940666666666664\\
1.11803693116646	0.950999999999998\\
1.14951905392452	0.960166666666665\\
1.19210578575966	0.969999999999999\\
1.23538561169301	0.980499999999999\\
1.29675576969867	0.9905\\
1.90373532593186	1\\
};

\plotRinitFive{matGray}{forget plot, only marks}
  table[row sep=crcr]{%
6.18488775898091e-06	0\\
0.000821022654975755	0.254166666666661\\
0.00162332794697027	0.49733333333331\\
0.00306616224214895	0.747166666666654\\
1.90373532593186	1\\
};

\end{axis}
\end{tikzpicture}%

%% file: CDF_Coop_RI.tex
%
%
\definecolor{mycolor1}{rgb}{0.85098,0.32549,0.09804}%
\begin{tikzpicture}
\begin{axis}[%
width=0.85\columnwidth,
height=0.3 \columnwidth,
scale only axis,
xmode=log,
xmin=1e-05,
xmax=2,
xminorticks=true,
xlabel={position error [m]},
ymin=0,
ymax=1,
ytick={  0,  0.25, 0.5, 0.75, 1},
ylabel={empirical CDF},
axis background/.style={fill=white},
xmajorgrids,
xminorgrids,
ymajorgrids,
grid style={gridGray},
legend style={at={(0.015,0.97)}, anchor=north west, legend cell align=left, align=left, draw=white!15!black, font=\scriptsize}
]

\plotPinit{black}{}
   table[row sep=crcr]{%
   100 1\\
   };
\addlegendentry{perf. init}

\plotPinit{black}{forget plot, no marks}
   table[row sep=crcr]{%
2.44404543152941e-06	0\\
2.70133863630007e-05	0.0101666666666664\\
4.18185789601653e-05	0.0214999999999994\\
5.72548893475743e-05	0.0309999999999999\\
6.89191280268638e-05	0.0408333333333332\\
8.15704324607093e-05	0.0504999999999997\\
9.20064175916725e-05	0.0604999999999996\\
0.000103149157966775	0.0701666666666668\\
0.000114270642413697	0.0798333333333338\\
0.000125502525913436	0.0893333333333339\\
0.000138776229061795	0.100000000000001\\
0.00015005178806412	0.1105\\
0.000158745680420015	0.121\\
0.000165637952848123	0.131333333333333\\
0.000177324874565293	0.141166666666666\\
0.000186890555436262	0.150999999999999\\
0.000196686706097555	0.160999999999998\\
0.000205943221794193	0.171166666666664\\
0.000213501010833779	0.180666666666664\\
0.000222069711106522	0.190499999999997\\
0.00023268546354803	0.200499999999996\\
0.00024148623055132	0.210333333333329\\
0.000249975045137991	0.220666666666662\\
0.000257563145037528	0.230499999999994\\
0.000265804721975366	0.240833333333327\\
0.000272916919410809	0.25066666666666\\
0.000280599902425952	0.260333333333326\\
0.000288983857564908	0.270166666666659\\
0.000300493913154716	0.279999999999992\\
0.000308127661558134	0.290166666666658\\
0.000315916513853251	0.300166666666657\\
0.000322568177722137	0.30999999999999\\
0.00033171473131283	0.320333333333323\\
0.000340061788951118	0.330666666666656\\
0.000351204626884484	0.341166666666655\\
0.000359346679333154	0.351833333333322\\
0.000366235478358151	0.361833333333321\\
0.000375023006326604	0.371499999999987\\
0.00038406399022712	0.381833333333319\\
0.000392512332765294	0.391999999999985\\
0.000403349160362261	0.402499999999984\\
0.000413240993964737	0.412333333333316\\
0.000422207970647659	0.422499999999982\\
0.000436201991835288	0.433833333333314\\
0.000443909292675015	0.443666666666647\\
0.000452259876659416	0.454166666666646\\
0.00046347953856615	0.463999999999978\\
0.000474658323168199	0.473999999999977\\
0.000484174922344429	0.483999999999976\\
0.000497245156975701	0.493666666666642\\
0.000507903901079091	0.504333333333309\\
0.000516761390227451	0.515499999999976\\
0.000527361513101339	0.525666666666642\\
0.000539410047695438	0.53633333333331\\
0.000549079453949469	0.54633333333331\\
0.000559006595098921	0.55683333333331\\
0.000570249559047808	0.566833333333311\\
0.000582804649931434	0.576833333333312\\
0.000593045891329289	0.587166666666645\\
0.000605555427314273	0.596833333333312\\
0.000617573359687506	0.60649999999998\\
0.000630096901802614	0.61699999999998\\
0.000644374064720225	0.626833333333314\\
0.000658149221930217	0.636499999999981\\
0.000671555394192521	0.646166666666648\\
0.000683241176494166	0.656999999999982\\
0.000697392181412746	0.666833333333315\\
0.000717784623718089	0.676666666666649\\
0.000732689080011819	0.686999999999983\\
0.000751010547812443	0.697833333333317\\
0.000773875535711089	0.70766666666665\\
0.000793532494024755	0.718166666666651\\
0.00081378613724347	0.728166666666652\\
0.000835605565038775	0.738333333333319\\
0.000853457875976975	0.748333333333319\\
0.000870928610223962	0.758166666666653\\
0.00089615629854489	0.768666666666654\\
0.000917809572160778	0.778666666666654\\
0.000946694381837467	0.788666666666655\\
0.000970054508496825	0.798333333333322\\
0.0010003606926027	0.808666666666656\\
0.00103049252791258	0.818333333333323\\
0.00106612571064881	0.827999999999991\\
0.00109792045813517	0.838499999999991\\
0.00113106234483239	0.848666666666658\\
0.00116368103408167	0.858999999999992\\
0.00119962404901172	0.868833333333326\\
0.00123890976740538	0.878999999999993\\
0.00128839295751701	0.889333333333327\\
0.0013411393713242	0.899166666666661\\
0.00139618083131708	0.909166666666662\\
0.00146723250120337	0.919166666666662\\
0.00153740707319706	0.929166666666663\\
0.00163160106427526	0.93933333333333\\
0.00173176940609292	0.949333333333331\\
0.00184898348930868	0.959499999999998\\
0.00203039512562831	0.969999999999999\\
0.00224582686620488	0.979999999999999\\
0.00278672300446018	0.990499999999999\\
0.00530482744873202	1\\
};

\plotPinit{black}{forget plot, only marks}
   table[row sep=crcr]{%
2.44404543152941e-06	0\\
0.000272916919410809	0.25066666666666\\
0.000507903901079091	0.504333333333309\\
0.000853457875976975	0.748333333333319\\
0.00530482744873202	1\\
};


\plotRinitOne{black}{}
  table[row sep=crcr]{%
100 1\\
};
\addlegendentry{1 rand. init}

\plotRinitOne{black}{forget plot, no marks}
  table[row sep=crcr]{%
6.09326018824486e-05	0\\
0.00303810868029213	0.00949999999999984\\
0.0105856291196316	0.0189999999999994\\
0.0248188729751556	0.0289999999999997\\
0.0447766885580838	0.0381666666666662\\
0.0651379049896061	0.0483333333333328\\
0.089556987668131	0.0584999999999992\\
0.115839593064492	0.0678333333333329\\
0.143452768538317	0.0779999999999997\\
0.165023335063804	0.0889999999999999\\
0.193040448018297	0.0993333333333327\\
0.217932938952422	0.109166666666666\\
0.246386959744457	0.118499999999999\\
0.281279244013112	0.128166666666665\\
0.305669258748047	0.137666666666665\\
0.337221550345001	0.146666666666665\\
0.363420536603312	0.156333333333332\\
0.398128556515156	0.165499999999998\\
0.422382219245815	0.176499999999998\\
0.45525697750878	0.186833333333331\\
0.480281995334001	0.197166666666664\\
0.50598527846079	0.206499999999996\\
0.533924282015751	0.216499999999996\\
0.563341738942713	0.225999999999995\\
0.58703821577819	0.235833333333328\\
0.616457896837469	0.245499999999994\\
0.634717941965872	0.254666666666661\\
0.653682394050962	0.264333333333327\\
0.675491312510949	0.274333333333326\\
0.701614079361396	0.284166666666659\\
0.725508720773821	0.294166666666659\\
0.742368787850346	0.304333333333325\\
0.765278752806427	0.314166666666657\\
0.787784736715206	0.324166666666657\\
0.80945073291077	0.333999999999989\\
0.831614530809059	0.343333333333322\\
0.847200658917193	0.353499999999988\\
0.866595646681956	0.362999999999987\\
0.88420506071788	0.374166666666654\\
0.904606233309439	0.384499999999987\\
0.922105673176308	0.394499999999986\\
0.941719877619429	0.404666666666652\\
0.964478692380894	0.415499999999984\\
0.981558495449069	0.42466666666665\\
0.998956299289312	0.435333333333316\\
1.01658969818075	0.445833333333315\\
1.03314577550935	0.455999999999981\\
1.04667890328564	0.466166666666647\\
1.06467535311682	0.476666666666646\\
1.08269894567718	0.486666666666645\\
1.09919131862724	0.497166666666645\\
1.11590944831155	0.506666666666645\\
1.12959971891567	0.516166666666645\\
1.14582646990328	0.525499999999979\\
1.15781271548493	0.536499999999979\\
1.17100756051615	0.546333333333313\\
1.18844624694048	0.55649999999998\\
1.20288210695872	0.566333333333314\\
1.2215848013682	0.575833333333314\\
1.23308986936662	0.584999999999981\\
1.24530395546839	0.595499999999982\\
1.25856460993406	0.604333333333315\\
1.27414410175996	0.615166666666649\\
1.28749806208525	0.624999999999983\\
1.30487445650491	0.634333333333317\\
1.31891974148107	0.644666666666651\\
1.33315066843098	0.654333333333318\\
1.34812646666521	0.663833333333318\\
1.36289627189082	0.673833333333318\\
1.3741783002703	0.683499999999985\\
1.3920432906559	0.694166666666652\\
1.40712164402284	0.704499999999986\\
1.41894958013045	0.713999999999986\\
1.43299078942148	0.72433333333332\\
1.44994026295171	0.734166666666654\\
1.46063108684571	0.744499999999988\\
1.47424965304445	0.754833333333321\\
1.49004800393723	0.764333333333322\\
1.50568153903711	0.775833333333323\\
1.51860925570689	0.787166666666657\\
1.53598457706546	0.796666666666657\\
1.55593689499574	0.806333333333324\\
1.5702356395216	0.816166666666658\\
1.58606413917876	0.826666666666659\\
1.6089680002878	0.836999999999993\\
1.62676345289013	0.846499999999993\\
1.65183197818913	0.856999999999994\\
1.66849412308213	0.868666666666661\\
1.69276714434928	0.879499999999995\\
1.71259078103812	0.889833333333329\\
1.74384392683978	0.900499999999996\\
1.76483970971941	0.912666666666663\\
1.79072343294584	0.923999999999997\\
1.81962277070719	0.933333333333331\\
1.86444677458408	0.944499999999998\\
1.89663846830064	0.954333333333332\\
1.95075953599063	0.964999999999999\\
2.0133502468094	0.975666666666666\\
2.08736090143102	0.986666666666666\\
2.39162353032851	1\\
};

\plotRinitOne{black}{forget plot, only marks}
  table[row sep=crcr]{%
6.09326018824486e-05	0\\
0.616457896837469	0.245499999999994\\
1.09919131862724	0.497166666666645\\
1.47424965304445	0.754833333333321\\
2.39162353032851	1\\
};


\plotRinitFive{black}{}
table[row sep=crcr]{%
100 1\\
};
\addlegendentry{5 rand. init}

\plotRinitFive{black}{forget plot, no marks}
table[row sep=crcr]{%
2.26897626531815e-05	0\\
0.000559437570903329	0.00966666666666649\\
0.00129999609604919	0.0196666666666662\\
0.00260928716154597	0.0298333333333329\\
0.00452222257226105	0.0401666666666666\\
0.00742449900354139	0.0506666666666666\\
0.0103496098594376	0.0606666666666668\\
0.0145234498497778	0.0703333333333339\\
0.0188587830766672	0.0811666666666676\\
0.0240633632118075	0.0903333333333343\\
0.0284398627511256	0.100333333333335\\
0.0341627462858566	0.109666666666667\\
0.0390156348527315	0.119166666666667\\
0.0456900330053902	0.128833333333333\\
0.0513809184537674	0.1395\\
0.0588801903949773	0.149333333333333\\
0.0659143196861823	0.158999999999999\\
0.0724766517928841	0.170833333333331\\
0.0792332300137808	0.180499999999998\\
0.0859502812382409	0.190166666666664\\
0.0921866764789323	0.199999999999996\\
0.0985517866839519	0.210499999999996\\
0.106714360247766	0.219999999999995\\
0.116092944047786	0.228999999999995\\
0.123118982130998	0.238333333333328\\
0.131573757169374	0.248166666666661\\
0.141765217120367	0.258333333333327\\
0.150736376732536	0.268499999999994\\
0.158770664019152	0.278499999999993\\
0.166316071913921	0.288833333333326\\
0.173303217598227	0.299166666666658\\
0.180720903508086	0.309166666666658\\
0.189711902463506	0.318833333333324\\
0.196398827943811	0.328666666666657\\
0.20505840123946	0.338166666666657\\
0.212576118337194	0.349166666666656\\
0.220502335414159	0.358666666666655\\
0.228932680228753	0.368999999999988\\
0.23996898070264	0.378666666666654\\
0.248147109623537	0.389999999999986\\
0.256925682261277	0.400333333333319\\
0.269561394428487	0.410666666666652\\
0.279902298954457	0.421166666666651\\
0.290013345236308	0.430833333333316\\
0.300331721256862	0.440333333333315\\
0.310046673750408	0.450333333333314\\
0.321320449608228	0.460833333333314\\
0.329223489793047	0.470166666666646\\
0.340745790789891	0.479999999999979\\
0.352022562620942	0.489499999999978\\
0.363737565856594	0.49933333333331\\
0.376671815963942	0.509166666666644\\
0.390231480109117	0.518999999999978\\
0.401745797789898	0.529333333333311\\
0.412998493298736	0.539333333333312\\
0.423493920622479	0.548999999999979\\
0.439364152269049	0.560166666666646\\
0.454721861566099	0.570666666666647\\
0.467759914557783	0.58149999999998\\
0.478478279135137	0.591499999999981\\
0.493081008999969	0.601333333333315\\
0.505960494308443	0.610999999999982\\
0.520091929205484	0.620833333333315\\
0.53417117157912	0.631499999999982\\
0.547834253028794	0.641833333333316\\
0.562728993552135	0.651999999999983\\
0.576720094504048	0.661333333333317\\
0.591904128363836	0.671499999999984\\
0.605247022837266	0.681499999999984\\
0.623739177171716	0.691333333333318\\
0.637544707876849	0.701666666666652\\
0.653098210869075	0.712166666666652\\
0.668473577675538	0.721999999999986\\
0.683632720795906	0.731166666666653\\
0.698396807962075	0.741999999999987\\
0.713019036182555	0.75183333333332\\
0.728257951351054	0.762166666666654\\
0.741969787429591	0.772499999999988\\
0.753565363438332	0.782833333333322\\
0.768536991183197	0.792666666666656\\
0.785680320392945	0.802166666666656\\
0.810868297144084	0.812166666666657\\
0.831197234344768	0.822333333333324\\
0.850095059020463	0.832833333333325\\
0.871713609948904	0.843999999999992\\
0.88728915969599	0.853999999999993\\
0.910249168736708	0.86516666666666\\
0.936203502351129	0.874999999999994\\
0.963640661952355	0.885833333333328\\
0.988307547835626	0.896499999999995\\
1.01498311528073	0.906499999999996\\
1.04407412747757	0.915833333333329\\
1.06986147010022	0.92583333333333\\
1.10242957814005	0.938166666666664\\
1.14502925021205	0.948333333333331\\
1.19021990347598	0.958166666666665\\
1.23734438869919	0.968999999999999\\
1.30653106159352	0.979166666666666\\
1.38605858939794	0.989\\
1.87001801262856	1\\
};

\plotRinitFive{black}{forget plot, only marks}
table[row sep=crcr]{%
2.26897626531815e-05	0\\
0.131573757169374	0.248166666666661\\
0.363737565856594	0.49933333333331\\
0.713019036182555	0.75183333333332\\
1.87001801262856	1\\
};

\end{axis}
\end{tikzpicture}%

%% file: CDF_3Step_Multilat.tex
%
%
%
\begin{tikzpicture}

\begin{axis}[%
width=0.85\columnwidth,
height=0.3 \columnwidth,
scale only axis,
xmode=log,
xmin=1e-05,
xmax=2,
xminorticks=true,
xlabel={position error [m]},
ymin=0,
ymax=1,
ytick={  0,  0.25, 0.5, 0.75, 1},
ylabel={empirical CDF},
axis background/.style={fill=white},
xmajorgrids,
xminorgrids,
ymajorgrids,
grid style={gridGray},
legend style={at={(1,0)}, anchor=south east, legend cell align=left, align=left, draw=white!15!black, font=\scriptsize}
]


\plotMultilat{matRed}{}
  table[row sep=crcr]{%
100	1\\
};
\addlegendentry{multilateration}

\plotMultilat{matRed}{forget plot, no marks}
  table[row sep=crcr]{%
0.00017705737275233	0\\
0.00144173967532729	0.0101666666666663\\
0.00187161272981873	0.020166666666666\\
0.00223407443304013	0.0301666666666662\\
0.00246731506459388	0.0399999999999997\\
0.00271833516813689	0.0498333333333331\\
0.00296786733558936	0.0606666666666665\\
0.00321438788893911	0.0710000000000002\\
0.00340071964156157	0.0808333333333338\\
0.00357042138156957	0.0905000000000005\\
0.0038054725988782	0.100833333333334\\
0.00398889356150896	0.111333333333333\\
0.00420419990626032	0.122\\
0.00444812888798253	0.132499999999999\\
0.00464285528169668	0.142333333333332\\
0.00484671810287823	0.152999999999999\\
0.00502503948239625	0.162666666666665\\
0.00523051077766993	0.172833333333331\\
0.00541110571996021	0.182666666666664\\
0.0055722217313741	0.192499999999996\\
0.00576443466178395	0.202499999999996\\
0.00593284835030539	0.212999999999995\\
0.00611395607891895	0.223166666666661\\
0.00631447055575125	0.234333333333327\\
0.00654628976416839	0.244833333333326\\
0.00680530608760764	0.255166666666659\\
0.00703353615345519	0.265333333333325\\
0.00722300206892449	0.275333333333325\\
0.00744096401786348	0.285333333333324\\
0.00764853505972155	0.295666666666657\\
0.0078743054181767	0.305333333333323\\
0.00803221362903889	0.315166666666656\\
0.00818947003355355	0.325333333333322\\
0.00841862820192516	0.335333333333322\\
0.00857170530995328	0.345666666666655\\
0.00873940563246636	0.355833333333321\\
0.00892092934956207	0.366166666666653\\
0.00918510385706024	0.376666666666652\\
0.00937030894445291	0.386666666666651\\
0.00959392790507378	0.396499999999984\\
0.00975862850053815	0.406499999999983\\
0.00996023014876839	0.416499999999982\\
0.0101769484346033	0.427166666666647\\
0.0103584087092134	0.43799999999998\\
0.0106053875190558	0.448333333333312\\
0.0108033201015951	0.458666666666644\\
0.0110274858742963	0.46883333333331\\
0.0112204897887289	0.478999999999976\\
0.0114044267349694	0.489499999999975\\
0.0116267196063075	0.499999999999974\\
0.0119361151998194	0.509833333333308\\
0.0121917892624483	0.519999999999975\\
0.0123941062524197	0.529666666666642\\
0.0125969214014561	0.539166666666642\\
0.0128350368522574	0.549666666666642\\
0.0130631005846207	0.559499999999976\\
0.0133024934874085	0.569999999999977\\
0.013639874282064	0.57983333333331\\
0.0138586448006658	0.589333333333311\\
0.0141933187122174	0.599166666666645\\
0.0145591085805904	0.608833333333312\\
0.014852650756654	0.619499999999979\\
0.0151760860694552	0.629333333333313\\
0.0154294008669523	0.63999999999998\\
0.0157164820761267	0.649833333333314\\
0.0160022831954474	0.659999999999981\\
0.016307202704792	0.669833333333315\\
0.0167053629518439	0.679999999999982\\
0.0169995721112618	0.690333333333315\\
0.0173357925643401	0.700499999999983\\
0.0177130059224719	0.710999999999983\\
0.0180576086759034	0.720833333333317\\
0.0183794112243329	0.730833333333317\\
0.0187498255233539	0.740333333333318\\
0.0190715626349175	0.750499999999985\\
0.019442523651712	0.760333333333319\\
0.0199251541962117	0.77083333333332\\
0.0204987559082775	0.780666666666654\\
0.0210541579332461	0.790999999999988\\
0.0215544344635261	0.800999999999988\\
0.0221476703377092	0.810999999999989\\
0.0227642128772105	0.820833333333323\\
0.0232726431062845	0.83049999999999\\
0.0238965525599755	0.840833333333324\\
0.0245580172390529	0.850666666666658\\
0.0250964922348443	0.860833333333325\\
0.0258759652449974	0.870499999999992\\
0.0268328489243383	0.880333333333326\\
0.027491231154074	0.890499999999993\\
0.0283440444019252	0.90016666666666\\
0.0292390434445655	0.910666666666661\\
0.0300895675870547	0.920333333333328\\
0.0318403581873892	0.930166666666662\\
0.033311527831028	0.940333333333329\\
0.0351789671091496	0.95033333333333\\
0.0377577530626193	0.960166666666664\\
0.0408404140481644	0.970333333333331\\
0.0454920313270002	0.979999999999999\\
0.0534195907839271	0.990333333333333\\
0.131382113239411	1\\
};

\plotMultilat{matRed}{forget plot, only marks}
  table[row sep=crcr]{%
0.00017705737275233	0\\
0.00654628976416839	0.244833333333326\\
0.0116267196063075	0.499999999999974\\
0.0190715626349175	0.750499999999985\\
0.131382113239411	1\\
};


\plotML{matBlue}{}
  table[row sep=crcr]{%
100	1\\
};
\addlegendentry{\pairML}

\plotML{matBlue}{forget plot, no marks}
  table[row sep=crcr]{%
3.10456375626894e-06	0\\
3.19534803057679e-05	0.0101666666666664\\
5.01169645569431e-05	0.0208333333333329\\
7.15224608698949e-05	0.0303333333333334\\
9.99408641216443e-05	0.0398333333333335\\
0.000135896096195714	0.0495\\
0.000172034180830825	0.059333333333333\\
0.000201470755106757	0.0690000000000004\\
0.000230736666976473	0.0786666666666673\\
0.000260494948713772	0.0883333333333342\\
0.000293149426962524	0.0983333333333339\\
0.000328654053865324	0.108\\
0.000358190573680591	0.1175\\
0.000390465860614856	0.127499999999999\\
0.000432146230868137	0.137499999999999\\
0.000473548733127574	0.147333333333332\\
0.000522144476049178	0.157333333333332\\
0.00057267134769887	0.166999999999998\\
0.000614533964296188	0.176999999999998\\
0.000652445420097415	0.186499999999997\\
0.000697854776472962	0.196166666666663\\
0.000739279487242052	0.206666666666663\\
0.000795964552756848	0.216666666666662\\
0.00084890609584595	0.226333333333328\\
0.000915061724455172	0.236166666666661\\
0.000973229509214245	0.245999999999994\\
0.00102226842436562	0.256333333333327\\
0.00107787719783444	0.266833333333326\\
0.00113118233742713	0.277833333333326\\
0.00117603352743318	0.287499999999992\\
0.00122621473655357	0.297166666666658\\
0.00129542498576231	0.307333333333324\\
0.00134679977481841	0.317333333333324\\
0.00140897864400053	0.32699999999999\\
0.00147456525596926	0.336999999999989\\
0.00156760287591811	0.347333333333322\\
0.00163832985648731	0.357166666666654\\
0.0017065810098528	0.366999999999987\\
0.00178338113793609	0.376666666666652\\
0.00186636268059196	0.386833333333318\\
0.00193083277570912	0.396666666666651\\
0.00200458716576706	0.407333333333316\\
0.00209179153664969	0.417333333333315\\
0.00216236515262158	0.427499999999981\\
0.00222550093723604	0.437333333333313\\
0.00232071567563092	0.447666666666646\\
0.00241501298597169	0.458499999999978\\
0.00249845559277784	0.468499999999977\\
0.00258258547333239	0.47883333333331\\
0.00264626158042747	0.488833333333309\\
0.00277161632442534	0.498833333333308\\
0.00287746103295175	0.509333333333309\\
0.0029723457611129	0.519333333333309\\
0.00306208277889148	0.529333333333309\\
0.00315173149172347	0.539333333333309\\
0.00322811712390254	0.549999999999976\\
0.00331240088089742	0.55983333333331\\
0.00342474044727112	0.570499999999978\\
0.00353058634848722	0.581166666666645\\
0.00364673084732636	0.591166666666645\\
0.00376408569584852	0.601499999999979\\
0.0038960038234173	0.611833333333313\\
0.00401665733060066	0.62199999999998\\
0.00413354438844749	0.632499999999981\\
0.00423011395215833	0.642999999999981\\
0.00437016800480837	0.652999999999982\\
0.00452715905616766	0.663666666666648\\
0.00467550028043987	0.673999999999982\\
0.00483119216754112	0.685166666666649\\
0.00495334352563011	0.69516666666665\\
0.00511545199090711	0.70516666666665\\
0.00525810128730777	0.715166666666651\\
0.00543178425247575	0.725999999999985\\
0.00568434733329653	0.736833333333319\\
0.00584806784796428	0.747166666666652\\
0.00604865340229486	0.757166666666653\\
0.00629588535901932	0.76783333333332\\
0.00658391152018945	0.777499999999987\\
0.00683687771008907	0.787166666666655\\
0.00710411111734156	0.796999999999988\\
0.00730027243232123	0.807499999999989\\
0.00752438622049491	0.817666666666656\\
0.0077674188782253	0.828166666666657\\
0.00801223497391935	0.838166666666658\\
0.00833674677499168	0.848333333333325\\
0.00872014224938841	0.858333333333325\\
0.00908769096487604	0.868499999999993\\
0.00943399027388849	0.879499999999993\\
0.00991704079663498	0.889499999999994\\
0.0103963029920629	0.899499999999994\\
0.0107480916463373	0.909333333333328\\
0.0114650578867079	0.919666666666662\\
0.0122389551938252	0.929833333333329\\
0.0128946288879253	0.940166666666663\\
0.014238525690719	0.950499999999997\\
0.0155421731723209	0.960166666666664\\
0.0170092343239992	0.970166666666665\\
0.0192284616730264	0.979666666666665\\
0.0225986604036476	0.990166666666666\\
0.0408287325134743	1\\
};

\plotML{matBlue}{forget plot, only marks}
  table[row sep=crcr]{%
3.10456375626894e-06	0\\
0.000973229509214245	0.245999999999994\\
0.00277161632442534	0.498833333333308\\
0.00584806784796428	0.747166666666652\\
0.0408287325134743	1\\
};


\plotAdvinit{black}{}
table[row sep=crcr]{%
100	1\\
};
\addlegendentry{\turboLS}

\plotAdvinit{matGray}{forget plot, no marks}
  table[row sep=crcr]{%
3.10231667553882e-06	0\\
3.27732244057998e-05	0.0101666666666663\\
5.01325582154658e-05	0.0208333333333328\\
7.24114564831451e-05	0.0303333333333333\\
9.89616023389334e-05	0.0398333333333334\\
0.000128353305496403	0.0496666666666664\\
0.000165598636255824	0.0598333333333331\\
0.000200178654141401	0.0695000000000003\\
0.000224305120491631	0.079333333333334\\
0.000251212698366192	0.0890000000000009\\
0.000279853156441485	0.0990000000000009\\
0.000310740601271972	0.108666666666667\\
0.00034078932397412	0.118333333333333\\
0.000364666667810446	0.128166666666666\\
0.000393598707511426	0.137833333333332\\
0.000424310088869912	0.147666666666666\\
0.000453522855266532	0.157166666666665\\
0.000491281033742039	0.167333333333332\\
0.00052858802037529	0.176999999999998\\
0.000560380797685423	0.186999999999997\\
0.000591246917781626	0.197666666666663\\
0.000618026043616854	0.208166666666663\\
0.000648209715868115	0.217999999999996\\
0.000677405238987987	0.227999999999995\\
0.000711109974825884	0.237999999999994\\
0.00074019531729322	0.24866666666666\\
0.00076515407373764	0.25816666666666\\
0.000796541878204189	0.268333333333326\\
0.00082208574395587	0.278499999999992\\
0.000855184163678593	0.288666666666658\\
0.000888571361687211	0.299666666666657\\
0.000918875770300729	0.310666666666657\\
0.000951938005921436	0.32099999999999\\
0.000991305771138373	0.330833333333323\\
0.00101703153949561	0.340999999999989\\
0.00104157952635319	0.351333333333322\\
0.00107296656181778	0.361666666666654\\
0.0011029993863611	0.372166666666654\\
0.00112773699430578	0.382166666666653\\
0.00115520841185022	0.392166666666651\\
0.00119145147494301	0.401833333333317\\
0.00122683426392642	0.411833333333316\\
0.00125497112041193	0.421499999999982\\
0.00129377196139191	0.431499999999981\\
0.00131790787462768	0.441333333333313\\
0.00134481630743833	0.451999999999979\\
0.0013727632572406	0.462499999999978\\
0.00140453793150183	0.47233333333331\\
0.00143361319164263	0.482166666666643\\
0.0014634867839556	0.491999999999975\\
0.0015069738313316	0.501666666666641\\
0.00153875015797262	0.511833333333308\\
0.00157367404184108	0.521999999999975\\
0.00161144306091863	0.531666666666642\\
0.00164238453984501	0.541333333333309\\
0.00167508827217045	0.551166666666643\\
0.00171289049459028	0.560999999999977\\
0.00175397422234447	0.571166666666644\\
0.00179669592977007	0.581166666666644\\
0.00183095496564912	0.591499999999978\\
0.00187320549948823	0.601499999999979\\
0.0019108594657205	0.611999999999979\\
0.00193705982051919	0.622166666666646\\
0.00198824896027145	0.631666666666647\\
0.0020323039484693	0.642166666666647\\
0.00207588422460413	0.651999999999981\\
0.00211874178817243	0.662666666666648\\
0.00215905870193	0.672833333333315\\
0.00221430362112586	0.683333333333315\\
0.00227121160062429	0.69416666666665\\
0.00231882216757419	0.704833333333317\\
0.0023666979411626	0.71466666666665\\
0.00241805285416056	0.724999999999985\\
0.00248586141909661	0.734999999999985\\
0.0025334339626953	0.745833333333319\\
0.00259069463487613	0.755999999999986\\
0.00265772156344684	0.76583333333332\\
0.00272888431528246	0.776166666666654\\
0.00280915033803773	0.786333333333321\\
0.00290516459025411	0.796833333333322\\
0.002996400703739	0.806999999999989\\
0.00305963080865595	0.817166666666656\\
0.00314073583692783	0.827166666666657\\
0.00323417248099243	0.837666666666658\\
0.00331817576568701	0.847833333333325\\
0.00342314605071209	0.858499999999992\\
0.00355977669434209	0.869166666666659\\
0.00370566298231448	0.879333333333327\\
0.00386942071880057	0.88966666666666\\
0.0040461643940413	0.899833333333328\\
0.00419001029068671	0.909833333333328\\
0.00439805002580222	0.919499999999996\\
0.00459620687795969	0.929499999999996\\
0.0048381492576728	0.939499999999997\\
0.00517066065470142	0.949666666666664\\
0.00556694614329632	0.959333333333331\\
0.00620250018860761	0.969833333333332\\
0.00684250481122101	0.979833333333332\\
0.00803421665725137	0.990333333333333\\
0.0157524157248099	1\\
};

\plotAdvinit{matGray}{forget plot, only marks}
  table[row sep=crcr]{%
3.10231667553882e-06	0\\
0.00074019531729322	0.24866666666666\\
0.0015069738313316	0.501666666666641\\
0.0025334339626953	0.745833333333319\\
0.0157524157248099	1\\
};


\plotAdvinit{black}{forget plot, no marks}
 table[row sep=crcr]{%
2.44404375903542e-06	0\\
2.70133857140138e-05	0.0101666666666664\\
4.18185788990934e-05	0.0214999999999994\\
5.72545036423179e-05	0.0309999999999999\\
6.89191588156905e-05	0.0408333333333332\\
8.15704298744232e-05	0.0504999999999997\\
9.20064159549584e-05	0.0604999999999996\\
0.000103149150374532	0.0701666666666668\\
0.000114270700181047	0.0798333333333338\\
0.00012550252181527	0.0893333333333339\\
0.000138776315961634	0.100000000000001\\
0.000150051812810258	0.1105\\
0.000158745713412296	0.121\\
0.000165687276174585	0.131333333333333\\
0.000177324892564651	0.141166666666666\\
0.000186890547008944	0.150999999999999\\
0.000196688301313312	0.160999999999998\\
0.000205980398435582	0.171166666666664\\
0.000213500997498	0.180666666666664\\
0.000222069657722391	0.190499999999997\\
0.000232684805367484	0.200499999999996\\
0.000241486252444256	0.210333333333329\\
0.000249975416030384	0.220666666666662\\
0.000257563260753596	0.230499999999994\\
0.000265804735423676	0.240833333333327\\
0.000272916995665552	0.25066666666666\\
0.000280599793051363	0.260333333333326\\
0.000288983943153488	0.270166666666659\\
0.0003004940317456	0.279999999999992\\
0.000308127374557343	0.290166666666658\\
0.00031591650362213	0.300166666666657\\
0.000322568149746381	0.30999999999999\\
0.000331714708252985	0.320333333333323\\
0.000340067309623661	0.330666666666656\\
0.000351204558983455	0.341166666666655\\
0.000359346698265309	0.351833333333322\\
0.000366235800200026	0.361833333333321\\
0.000375023519723639	0.371499999999987\\
0.000384063965899683	0.381833333333319\\
0.000392512385442998	0.391999999999985\\
0.000403349093846302	0.402499999999984\\
0.000413241043526124	0.412333333333316\\
0.000422208021557398	0.422499999999982\\
0.00043620203574146	0.433833333333314\\
0.000443909309899386	0.443666666666647\\
0.000452258681923415	0.454166666666646\\
0.000463479561383222	0.463999999999978\\
0.000474658301856783	0.473999999999977\\
0.000484174922273757	0.483999999999976\\
0.000497245955758182	0.493666666666642\\
0.000507903658864138	0.504333333333309\\
0.000516761413938669	0.515499999999976\\
0.000527361512069011	0.525666666666642\\
0.000539408589461849	0.53633333333331\\
0.000549079513819544	0.54633333333331\\
0.000559006662275472	0.55683333333331\\
0.000570249543186956	0.566833333333311\\
0.000582804599793891	0.576833333333312\\
0.000593045843650563	0.587166666666645\\
0.000605555425580197	0.596833333333312\\
0.000617572104450575	0.60649999999998\\
0.00062987122250732	0.61699999999998\\
0.000644374178317829	0.626833333333314\\
0.000658149448703486	0.636499999999981\\
0.00067155402522948	0.646166666666648\\
0.000683243954228968	0.656999999999982\\
0.00069738948656733	0.666833333333315\\
0.000717784688242984	0.676666666666649\\
0.000732689236974446	0.686999999999983\\
0.000751010636066804	0.697833333333317\\
0.000773875669494904	0.70766666666665\\
0.000793532495734682	0.718166666666651\\
0.000813785415872562	0.728166666666652\\
0.000835605585571267	0.738333333333319\\
0.000853459700004226	0.748333333333319\\
0.000870928560529506	0.758166666666653\\
0.000896156052802367	0.768666666666654\\
0.000917809606417644	0.778666666666654\\
0.000946694327443878	0.788666666666655\\
0.000970054705192312	0.798333333333322\\
0.00100036082625654	0.808666666666656\\
0.00103049347331049	0.818333333333323\\
0.00106612569092282	0.827999999999991\\
0.0010979207804997	0.838499999999991\\
0.00113106035176016	0.848666666666658\\
0.0011636806868265	0.858999999999992\\
0.00119962401923018	0.868833333333326\\
0.00123890995270928	0.878999999999993\\
0.00128839237224511	0.889333333333327\\
0.0013411399121947	0.899166666666661\\
0.00139459625231418	0.909166666666662\\
0.0014672330068329	0.919166666666662\\
0.0015374075193514	0.929166666666663\\
0.00163160259761694	0.93933333333333\\
0.00173176894855611	0.949333333333331\\
0.00184898353814502	0.959499999999998\\
0.00203042478672694	0.969999999999999\\
0.00224582612566905	0.979999999999999\\
0.00278672302490038	0.990499999999999\\
0.00530482758213705	1\\
};

\plotAdvinit{black}{forget plot, only marks}
 table[row sep=crcr]{%
2.44404375903542e-06	0\\
0.000272916995665552	0.25066666666666\\
0.000507903658864138	0.504333333333309\\
0.000853459700004226	0.748333333333319\\
0.00530482758213705	1\\
};


\plotPinit{black}{}
table[row sep=crcr]{%
100	1\\
};
\addlegendentry{\numLS, perf. init.}

\plotPinit{matGray}{forget plot, mark=none}
table[row sep=crcr]{%
3.2842863438936e-06	0\\
3.27734044777462e-05	0.0101666666666663\\
4.99809183392531e-05	0.0208333333333327\\
7.15215178632704e-05	0.0303333333333332\\
9.86656212658874e-05	0.0398333333333333\\
0.000128149246980109	0.0496666666666664\\
0.000165011138421207	0.0596666666666662\\
0.00019693745608415	0.0695\\
0.000223430479058421	0.0793333333333337\\
0.000250367455665966	0.0890000000000007\\
0.000277898643862753	0.0990000000000008\\
0.000309419670925857	0.108833333333333\\
0.000341014412939451	0.118333333333333\\
0.000365749342727433	0.128166666666666\\
0.000393598466421361	0.137833333333333\\
0.000423718850766218	0.147666666666666\\
0.000457283564647887	0.157166666666666\\
0.000491281003737809	0.167333333333332\\
0.000531225254289406	0.176999999999998\\
0.00056074065829462	0.186999999999997\\
0.000591509784587231	0.197666666666663\\
0.000619076509069033	0.208166666666663\\
0.000648210475086847	0.217999999999995\\
0.000677404167757512	0.227999999999995\\
0.000710808342628737	0.237999999999994\\
0.000740195338155888	0.24866666666666\\
0.000764978705469697	0.258499999999993\\
0.000796436022380328	0.26866666666666\\
0.000820540970640621	0.278833333333326\\
0.000854840694668714	0.289166666666659\\
0.000885750455150127	0.300166666666658\\
0.000917145025008286	0.311166666666657\\
0.000950355509758901	0.321333333333323\\
0.000990943240354239	0.331333333333322\\
0.00101543483617363	0.341499999999988\\
0.00104039546744355	0.351666666666655\\
0.00107012719686915	0.361999999999987\\
0.00110299920102573	0.372166666666653\\
0.00112773689687534	0.382166666666652\\
0.00115520884473625	0.392166666666651\\
0.00119145127718432	0.401833333333317\\
0.00122683429599596	0.411833333333316\\
0.00125497124891152	0.421499999999982\\
0.00129377211849664	0.431499999999981\\
0.00131790888436846	0.441333333333313\\
0.00134481597653174	0.451999999999979\\
0.00137276317086137	0.462499999999978\\
0.00140454007366901	0.47233333333331\\
0.00143361302426257	0.482166666666643\\
0.00146348329999766	0.491999999999975\\
0.00150697414514773	0.501666666666641\\
0.0015378031619711	0.511833333333308\\
0.00157257921141179	0.521999999999975\\
0.00161113480277512	0.531666666666642\\
0.00164199368005524	0.541333333333309\\
0.00167355512004456	0.551166666666643\\
0.00171230486713249	0.560999999999977\\
0.00175397417523507	0.571166666666644\\
0.00179442389960523	0.581166666666644\\
0.00183050483354576	0.591499999999978\\
0.00187286709235398	0.601499999999979\\
0.00191066468761937	0.611999999999979\\
0.00193655085951919	0.622166666666646\\
0.00198785587277208	0.631666666666647\\
0.00203180300313122	0.642166666666648\\
0.00207588508230415	0.651999999999981\\
0.00211879380318111	0.662666666666648\\
0.00215905872419039	0.672833333333315\\
0.00221430257652085	0.683333333333315\\
0.00227120945979942	0.69416666666665\\
0.00231882198961029	0.704833333333317\\
0.00236669445610214	0.71466666666665\\
0.00241804807544971	0.724999999999985\\
0.00248585835647708	0.734999999999985\\
0.00253343399557336	0.745833333333319\\
0.00259069293028403	0.755999999999986\\
0.00265772154405582	0.76583333333332\\
0.0027288846679025	0.776166666666654\\
0.0028095021561273	0.786333333333321\\
0.00290515323427043	0.796833333333322\\
0.00299640068689915	0.806999999999989\\
0.00305963077036763	0.817166666666656\\
0.00314073547273102	0.827166666666657\\
0.00323417345337976	0.837666666666657\\
0.00331817576709091	0.847833333333325\\
0.00342314601348122	0.858499999999992\\
0.00355977748301921	0.869166666666659\\
0.0037056658994467	0.879333333333326\\
0.00386942108878251	0.88966666666666\\
0.00404616477689709	0.899833333333328\\
0.00419000963475914	0.909833333333328\\
0.00439804985629735	0.919499999999995\\
0.00459620800139127	0.929499999999996\\
0.00483814149012222	0.939499999999997\\
0.0051706567460741	0.949666666666664\\
0.00556694732504295	0.959333333333331\\
0.00620248978255262	0.969833333333332\\
0.00684250352659337	0.979833333333332\\
0.00803425670527542	0.990333333333333\\
0.0157524026609455	1\\
};

\plotPinit{matGray}{forget plot, only marks}
table[row sep=crcr]{%
3.2842863438936e-06	0\\
0.000740195338155888	0.24866666666666\\
0.00150697414514773	0.501666666666641\\
0.00253343399557336	0.745833333333319\\
0.0157524026609455	1\\
};


\plotPinit{black}{forget plot, no marks}
   table[row sep=crcr]{%
2.44404543152941e-06	0\\
2.70133863630007e-05	0.0101666666666664\\
4.18185789601653e-05	0.0214999999999994\\
5.72548893475743e-05	0.0309999999999999\\
6.89191280268638e-05	0.0408333333333332\\
8.15704324607093e-05	0.0504999999999997\\
9.20064175916725e-05	0.0604999999999996\\
0.000103149157966775	0.0701666666666668\\
0.000114270642413697	0.0798333333333338\\
0.000125502525913436	0.0893333333333339\\
0.000138776229061795	0.100000000000001\\
0.00015005178806412	0.1105\\
0.000158745680420015	0.121\\
0.000165637952848123	0.131333333333333\\
0.000177324874565293	0.141166666666666\\
0.000186890555436262	0.150999999999999\\
0.000196686706097555	0.160999999999998\\
0.000205943221794193	0.171166666666664\\
0.000213501010833779	0.180666666666664\\
0.000222069711106522	0.190499999999997\\
0.00023268546354803	0.200499999999996\\
0.00024148623055132	0.210333333333329\\
0.000249975045137991	0.220666666666662\\
0.000257563145037528	0.230499999999994\\
0.000265804721975366	0.240833333333327\\
0.000272916919410809	0.25066666666666\\
0.000280599902425952	0.260333333333326\\
0.000288983857564908	0.270166666666659\\
0.000300493913154716	0.279999999999992\\
0.000308127661558134	0.290166666666658\\
0.000315916513853251	0.300166666666657\\
0.000322568177722137	0.30999999999999\\
0.00033171473131283	0.320333333333323\\
0.000340061788951118	0.330666666666656\\
0.000351204626884484	0.341166666666655\\
0.000359346679333154	0.351833333333322\\
0.000366235478358151	0.361833333333321\\
0.000375023006326604	0.371499999999987\\
0.00038406399022712	0.381833333333319\\
0.000392512332765294	0.391999999999985\\
0.000403349160362261	0.402499999999984\\
0.000413240993964737	0.412333333333316\\
0.000422207970647659	0.422499999999982\\
0.000436201991835288	0.433833333333314\\
0.000443909292675015	0.443666666666647\\
0.000452259876659416	0.454166666666646\\
0.00046347953856615	0.463999999999978\\
0.000474658323168199	0.473999999999977\\
0.000484174922344429	0.483999999999976\\
0.000497245156975701	0.493666666666642\\
0.000507903901079091	0.504333333333309\\
0.000516761390227451	0.515499999999976\\
0.000527361513101339	0.525666666666642\\
0.000539410047695438	0.53633333333331\\
0.000549079453949469	0.54633333333331\\
0.000559006595098921	0.55683333333331\\
0.000570249559047808	0.566833333333311\\
0.000582804649931434	0.576833333333312\\
0.000593045891329289	0.587166666666645\\
0.000605555427314273	0.596833333333312\\
0.000617573359687506	0.60649999999998\\
0.000630096901802614	0.61699999999998\\
0.000644374064720225	0.626833333333314\\
0.000658149221930217	0.636499999999981\\
0.000671555394192521	0.646166666666648\\
0.000683241176494166	0.656999999999982\\
0.000697392181412746	0.666833333333315\\
0.000717784623718089	0.676666666666649\\
0.000732689080011819	0.686999999999983\\
0.000751010547812443	0.697833333333317\\
0.000773875535711089	0.70766666666665\\
0.000793532494024755	0.718166666666651\\
0.00081378613724347	0.728166666666652\\
0.000835605565038775	0.738333333333319\\
0.000853457875976975	0.748333333333319\\
0.000870928610223962	0.758166666666653\\
0.00089615629854489	0.768666666666654\\
0.000917809572160778	0.778666666666654\\
0.000946694381837467	0.788666666666655\\
0.000970054508496825	0.798333333333322\\
0.0010003606926027	0.808666666666656\\
0.00103049252791258	0.818333333333323\\
0.00106612571064881	0.827999999999991\\
0.00109792045813517	0.838499999999991\\
0.00113106234483239	0.848666666666658\\
0.00116368103408167	0.858999999999992\\
0.00119962404901172	0.868833333333326\\
0.00123890976740538	0.878999999999993\\
0.00128839295751701	0.889333333333327\\
0.0013411393713242	0.899166666666661\\
0.00139618083131708	0.909166666666662\\
0.00146723250120337	0.919166666666662\\
0.00153740707319706	0.929166666666663\\
0.00163160106427526	0.93933333333333\\
0.00173176940609292	0.949333333333331\\
0.00184898348930868	0.959499999999998\\
0.00203039512562831	0.969999999999999\\
0.00224582686620488	0.979999999999999\\
0.00278672300446018	0.990499999999999\\
0.00530482744873202	1\\
};

\plotPinit{black}{forget plot, only marks}
   table[row sep=crcr]{%
2.44404543152941e-06	0\\
0.000272916919410809	0.25066666666666\\
0.000507903901079091	0.504333333333309\\
0.000853457875976975	0.748333333333319\\
0.00530482744873202	1\\
};



\node[align=center, font=\color{matGray}] at (0.00019,0.85) {\scriptsize \NC};
\node[align=center, font=\color{black}] at (0.00004,0.3) {\scriptsize cooperative};
\draw [-latex, color=matGray](0.0007,0.85) -- (0.002,0.75);
\draw [-latex, color=black](0.0001,0.3) -- (0.00023,0.27);
\end{axis}

\end{tikzpicture}%

%% file: CDF_time.tex
%
%
\begin{tikzpicture}

\begin{axis}[%
width=0.8\columnwidth,
height=0.3 \columnwidth,
scale only axis,
xmode=log,
xmin=0.01,
xmax=100,
xminorticks=true,
xlabel={computation time [s]},
ymin=0.0122699386503067,
ymax=1.01226993865031,
ytick={  0,  0.25, 0.5, 0.75, 1},
ylabel={empirical CDF},
axis background/.style={fill=white},
xmajorgrids,
xminorgrids,
ymajorgrids,
grid style={gridGray},
legend style={at={(0.5,1.4)}, anchor=north, legend columns=2, legend cell align=left, align=left, draw=white!15!black, font=\scriptsize}
]


\plotMultilat{matRed}{}
  table[row sep=crcr]{%
200	1\\
};
\addlegendentry{multilateration}

\plotMultilat{matRed}{forget plot, no marks}
  table[row sep=crcr]{%
0.020718	0\\
0.020761	0.01\\
0.020797	0.02\\
0.02085	0.03\\
0.020889	0.04\\
0.020948	0.055\\
0.020996	0.0650000000000001\\
0.021039	0.0750000000000001\\
0.021073	0.085\\
0.021102	0.095\\
0.021169	0.105\\
0.02118	0.115\\
0.021223	0.125\\
0.021239	0.135\\
0.021276	0.145\\
0.021296	0.155\\
0.021332	0.165\\
0.02135	0.175\\
0.021357	0.185\\
0.021403	0.195\\
0.021423	0.205\\
0.021442	0.215\\
0.021467	0.225\\
0.021494	0.235\\
0.021518	0.245\\
0.021534	0.25\\
0.021537	0.26\\
0.021553	0.27\\
0.021593	0.28\\
0.021615	0.29\\
0.021659	0.3\\
0.021668	0.31\\
0.0217	0.32\\
0.021735	0.33\\
0.021746	0.34\\
0.021753	0.35\\
0.021767	0.36\\
0.021787	0.37\\
0.021803	0.38\\
0.021823	0.39\\
0.021853	0.4\\
0.021856	0.41\\
0.021863	0.42\\
0.021873	0.43\\
0.021878	0.44\\
0.021898	0.45\\
0.02191	0.46\\
0.021924	0.47\\
0.021964	0.48\\
0.021968	0.49\\
0.021982	0.5\\
0.022024	0.51\\
0.022044	0.52\\
0.022048	0.53\\
0.022054	0.54\\
0.022063	0.555\\
0.022126	0.565\\
0.022147	0.58\\
0.022173	0.59\\
0.022196	0.6\\
0.022227	0.61\\
0.022243	0.62\\
0.022257	0.63\\
0.02226	0.64\\
0.022263	0.65\\
0.02228	0.66\\
0.022292	0.67\\
0.022314	0.68\\
0.022356	0.695\\
0.0224	0.705\\
0.022457	0.715\\
0.022494	0.725\\
0.022501	0.735\\
0.022536	0.745\\
0.022576	0.755\\
0.022577	0.76\\
0.022622	0.77\\
0.022647	0.78\\
0.022661	0.79\\
0.022694	0.8\\
0.022728	0.81\\
0.022752	0.82\\
0.022787	0.83\\
0.022824	0.84\\
0.022881	0.85\\
0.022898	0.86\\
0.022938	0.87\\
0.023017	0.88\\
0.02315	0.89\\
0.023183	0.9\\
0.023188	0.91\\
0.023231	0.92\\
0.023528	0.93\\
0.023547	0.94\\
0.023643	0.95\\
0.024007	0.96\\
0.024364	0.97\\
0.024601	0.98\\
0.025149	0.99\\
0.17159	1\\
};

\plotMultilat{matRed}{forget plot, only marks}
  table[row sep=crcr]{%
0.020718	0\\
0.021534	0.25\\
0.021982	0.5\\
0.022536	0.745\\
0.17159	1\\
};


\plotAdvinit{black}{}
table[row sep=crcr]{%
200	1\\
};
\addlegendentry{\turboLS}

\plotAdvinit{matGray}{forget plot, no marks}
  table[row sep=crcr]{%
0.104239	0\\
0.117	0.01\\
0.117816	0.02\\
0.118414	0.03\\
0.118691	0.04\\
0.119228	0.05\\
0.120516	0.0600000000000001\\
0.120887	0.0700000000000001\\
0.122306	0.08\\
0.123198	0.09\\
0.123692	0.1\\
0.123936	0.11\\
0.124869	0.12\\
0.125656	0.13\\
0.126219	0.14\\
0.12669	0.15\\
0.126763	0.16\\
0.12702	0.17\\
0.127544	0.18\\
0.128338	0.19\\
0.128482	0.2\\
0.12883	0.21\\
0.129267	0.22\\
0.129571	0.23\\
0.129822	0.24\\
0.129988	0.255\\
0.130482	0.265\\
0.130894	0.275\\
0.131099	0.285\\
0.131264	0.295\\
0.131457	0.305\\
0.131858	0.315\\
0.132089	0.325\\
0.132506	0.335\\
0.132607	0.345\\
0.132817	0.355\\
0.132991	0.365\\
0.133071	0.375\\
0.133256	0.385\\
0.133465	0.395\\
0.133543	0.405\\
0.133887	0.415\\
0.134125	0.425\\
0.134419	0.435\\
0.134499	0.445\\
0.134597	0.455\\
0.134966	0.465\\
0.13518	0.475\\
0.135446	0.485\\
0.135502	0.495\\
0.135997	0.505\\
0.136276	0.515\\
0.136528	0.525\\
0.13662	0.535\\
0.136698	0.545\\
0.137167	0.555\\
0.137328	0.565\\
0.137471	0.575\\
0.137686	0.585\\
0.137912	0.595\\
0.138297	0.605\\
0.138472	0.615\\
0.138524	0.625\\
0.138581	0.635\\
0.139016	0.645\\
0.139152	0.655\\
0.139283	0.665\\
0.139364	0.675\\
0.140174	0.685\\
0.140436	0.695\\
0.140564	0.705\\
0.140602	0.715\\
0.140876	0.725\\
0.140947	0.735\\
0.14124	0.745\\
0.141398	0.76\\
0.141507	0.77\\
0.142204	0.78\\
0.142413	0.79\\
0.14265	0.8\\
0.142827	0.81\\
0.143199	0.82\\
0.143306	0.83\\
0.143473	0.84\\
0.143743	0.85\\
0.143836	0.86\\
0.144227	0.87\\
0.145782	0.88\\
0.146839	0.89\\
0.147086	0.9\\
0.147787	0.91\\
0.150075	0.92\\
0.151576	0.93\\
0.152753	0.94\\
0.153557	0.95\\
0.154457	0.96\\
0.155901	0.97\\
0.157291	0.98\\
0.160244	0.99\\
0.205294	1\\
};

\plotAdvinit{matGray}{forget plot, only marks}
  table[row sep=crcr]{%
0.104239	0\\
0.129988	0.255\\
0.135502	0.495\\
0.14124	0.745\\
0.205294	1\\
};


\plotAdvinit{black}{forget plot, no marks}
 table[row sep=crcr]{%
3.596523	0\\
3.600834	0.01\\
3.607722	0.02\\
3.629724	0.03\\
3.705799	0.04\\
3.925249	0.05\\
4.004328	0.0600000000000001\\
4.063749	0.0700000000000001\\
4.314303	0.08\\
4.394651	0.09\\
4.421004	0.1\\
4.458554	0.11\\
4.461538	0.12\\
4.523476	0.13\\
4.537509	0.14\\
4.71558	0.15\\
4.752037	0.16\\
4.800374	0.17\\
4.863336	0.18\\
5.100267	0.19\\
5.120874	0.2\\
5.182435	0.21\\
5.190592	0.22\\
5.197794	0.23\\
5.204829	0.24\\
5.269831	0.255\\
5.282996	0.265\\
5.319783	0.275\\
5.385939	0.285\\
5.50456	0.295\\
5.567608	0.305\\
5.590499	0.315\\
5.593019	0.325\\
5.602156	0.335\\
5.648199	0.345\\
5.67307	0.355\\
5.68082	0.365\\
5.734645	0.375\\
5.802888	0.385\\
5.910912	0.395\\
5.969617	0.405\\
5.975393	0.415\\
6.015653	0.425\\
6.078028	0.435\\
6.101873	0.445\\
6.245507	0.455\\
6.297748	0.465\\
6.368559	0.475\\
6.371438	0.485\\
6.374402	0.495\\
6.3767	0.505\\
6.429087	0.515\\
6.467184	0.525\\
6.49308	0.535\\
6.670616	0.545\\
6.693071	0.555\\
6.745058	0.565\\
6.772664	0.575\\
6.779046	0.585\\
6.845823	0.595\\
6.86968	0.605\\
6.988336	0.615\\
7.066504	0.625\\
7.155298	0.635\\
7.222395	0.645\\
7.326757	0.655\\
7.3518	0.665\\
7.535803	0.675\\
7.570237	0.685\\
7.598109	0.695\\
7.678917	0.705\\
7.761291	0.715\\
7.808649	0.725\\
7.814388	0.735\\
7.941685	0.745\\
7.983998	0.76\\
8.055398	0.77\\
8.238019	0.78\\
8.344544	0.79\\
8.36205	0.8\\
8.515856	0.81\\
8.777083	0.82\\
9.013338	0.83\\
9.163434	0.84\\
9.432658	0.85\\
9.709415	0.86\\
9.853732	0.87\\
9.929068	0.88\\
10.254051	0.89\\
10.464255	0.9\\
10.625068	0.91\\
10.679772	0.92\\
11.01155	0.93\\
11.745601	0.94\\
13.011876	0.95\\
13.778288	0.96\\
14.323651	0.97\\
15.447484	0.98\\
16.547378	0.99\\
19.865497	1\\
};

\plotAdvinit{black}{forget plot, only marks}
 table[row sep=crcr]{%
3.596523	0\\
5.269831	0.255\\
6.374402	0.495\\
7.941685	0.745\\
19.865497	1\\
};


\plotML{matBlue}{}
  table[row sep=crcr]{%
200	1\\
};
\addlegendentry{\pairML}

\plotML{matBlue}{forget plot, no marks}
  table[row sep=crcr]{%
0.013555	0\\
0.013563	0.01\\
0.013601	0.02\\
0.013603	0.025\\
0.013608	0.035\\
0.013626	0.05\\
0.013634	0.0700000000000001\\
0.01364	0.08\\
0.013644	0.09\\
0.013646	0.095\\
0.013648	0.105\\
0.013666	0.115\\
0.013693	0.13\\
0.013706	0.14\\
0.013707	0.145\\
0.013746	0.16\\
0.013751	0.17\\
0.013757	0.18\\
0.013765	0.19\\
0.013774	0.2\\
0.013784	0.205\\
0.01379	0.215\\
0.013798	0.225\\
0.013814	0.24\\
0.013818	0.255\\
0.01382	0.26\\
0.013824	0.275\\
0.013828	0.285\\
0.013839	0.295\\
0.013845	0.305\\
0.013864	0.315\\
0.013871	0.32\\
0.013892	0.34\\
0.013901	0.355\\
0.013931	0.365\\
0.013949	0.375\\
0.013952	0.38\\
0.013957	0.39\\
0.01396	0.4\\
0.01399	0.41\\
0.013998	0.425\\
0.014001	0.44\\
0.014007	0.445\\
0.014012	0.455\\
0.01402	0.47\\
0.014029	0.485\\
0.014035	0.495\\
0.014036	0.505\\
0.014041	0.515\\
0.014066	0.525\\
0.014082	0.54\\
0.01409	0.555\\
0.014099	0.57\\
0.014102	0.575\\
0.014109	0.585\\
0.014118	0.595\\
0.014121	0.61\\
0.014128	0.62\\
0.014129	0.625\\
0.01414	0.635\\
0.01415	0.645\\
0.014152	0.655\\
0.014165	0.665\\
0.01418	0.675\\
0.014184	0.68\\
0.014209	0.69\\
0.014218	0.7\\
0.014225	0.71\\
0.014239	0.72\\
0.014241	0.725\\
0.014249	0.735\\
0.014267	0.745\\
0.014275	0.755\\
0.01428	0.765\\
0.014309	0.775\\
0.014314	0.78\\
0.014382	0.79\\
0.014402	0.8\\
0.014423	0.81\\
0.014503	0.82\\
0.014525	0.825\\
0.014567	0.835\\
0.014606	0.845\\
0.014652	0.855\\
0.014658	0.865\\
0.014674	0.875\\
0.014675	0.88\\
0.014683	0.89\\
0.014832	0.9\\
0.014845	0.91\\
0.014955	0.92\\
0.014979	0.925\\
0.015097	0.935\\
0.01516	0.945\\
0.015379	0.955\\
0.015646	0.965\\
0.016254	0.975\\
0.016341	0.98\\
0.017935	0.99\\
0.085587	1\\
};

\plotML{matBlue}{forget plot, only marks}
  table[row sep=crcr]{%
0.013555	0\\
0.013818	0.255\\
0.014035	0.495\\
0.014267	0.745\\
0.085587	1\\
};


\plotRinitOne{black}{}
table[row sep=crcr]{%
200	1\\
};
\addlegendentry{\numLS, 1 rand. init.}

\plotRinitOne{matGray}{forget plot, mark=none}
table[row sep=crcr]{%
0.535153	0\\
0.548743	0.01\\
0.599026	0.02\\
0.623279	0.03\\
0.625786	0.04\\
0.635963	0.05\\
0.641104	0.0600000000000001\\
0.66002	0.0700000000000001\\
0.666385	0.08\\
0.668341	0.09\\
0.682227	0.1\\
0.695079	0.11\\
0.708087	0.12\\
0.72181	0.13\\
0.72794	0.14\\
0.738117	0.15\\
0.746099	0.16\\
0.748814	0.17\\
0.75148	0.18\\
0.755241	0.19\\
0.760568	0.2\\
0.761555	0.21\\
0.76784	0.22\\
0.776279	0.23\\
0.785175	0.24\\
0.798122	0.255\\
0.80238	0.265\\
0.810731	0.275\\
0.813224	0.285\\
0.82047	0.295\\
0.825093	0.305\\
0.833694	0.315\\
0.835933	0.325\\
0.839867	0.335\\
0.840427	0.345\\
0.842363	0.355\\
0.84659	0.365\\
0.848074	0.375\\
0.849534	0.385\\
0.855814	0.395\\
0.86027	0.405\\
0.864326	0.415\\
0.867551	0.425\\
0.876686	0.435\\
0.879427	0.445\\
0.880394	0.455\\
0.882847	0.465\\
0.885143	0.475\\
0.890633	0.485\\
0.892941	0.495\\
0.897309	0.505\\
0.907778	0.515\\
0.912623	0.525\\
0.930311	0.535\\
0.948915	0.545\\
0.951396	0.555\\
0.956509	0.565\\
0.970119	0.575\\
0.971961	0.585\\
0.9759	0.595\\
0.987981	0.605\\
0.993075	0.615\\
0.993899	0.625\\
1.001813	0.635\\
1.006027	0.645\\
1.006646	0.655\\
1.016176	0.665\\
1.022088	0.675\\
1.025295	0.685\\
1.027502	0.695\\
1.057113	0.705\\
1.070043	0.715\\
1.083453	0.725\\
1.099528	0.735\\
1.100824	0.745\\
1.112904	0.76\\
1.130826	0.77\\
1.135283	0.78\\
1.151934	0.79\\
1.153595	0.8\\
1.157619	0.81\\
1.17255	0.82\\
1.194028	0.83\\
1.208699	0.84\\
1.219034	0.85\\
1.224398	0.86\\
1.23326	0.87\\
1.240914	0.88\\
1.25077	0.89\\
1.256114	0.9\\
1.283716	0.91\\
1.313003	0.92\\
1.365576	0.93\\
1.414907	0.94\\
1.521549	0.95\\
1.563411	0.96\\
1.66574	0.97\\
1.758859	0.98\\
1.80951	0.99\\
1.984414	1\\
};

\plotRinitOne{matGray}{forget plot, only marks}
table[row sep=crcr]{%
0.535153	0\\
0.798122	0.255\\
0.892941	0.495\\
1.100824	0.745\\
1.984414	1\\
};


\plotRinitOne{black}{forget plot, no marks}
   table[row sep=crcr]{%
54.938278	0\\
62.853234	0.01\\
63.225553	0.02\\
63.374378	0.03\\
63.413156	0.04\\
63.497974	0.05\\
63.578078	0.0600000000000001\\
63.587157	0.0700000000000001\\
63.694928	0.08\\
63.714304	0.09\\
63.734545	0.1\\
63.746597	0.11\\
63.777883	0.12\\
63.803929	0.13\\
63.818437	0.14\\
63.842871	0.15\\
63.868244	0.16\\
63.876474	0.17\\
63.944571	0.18\\
64.045219	0.19\\
64.092533	0.2\\
64.195014	0.21\\
64.328275	0.22\\
64.397378	0.23\\
64.424988	0.24\\
64.456802	0.255\\
64.490444	0.265\\
64.506996	0.275\\
64.528508	0.285\\
64.533721	0.295\\
64.541012	0.305\\
64.551239	0.315\\
64.557247	0.325\\
64.559534	0.335\\
64.561781	0.345\\
64.573919	0.355\\
64.597542	0.365\\
64.609447	0.375\\
64.625355	0.385\\
64.634668	0.395\\
64.646286	0.405\\
64.671743	0.415\\
64.678175	0.425\\
64.694042	0.435\\
64.713862	0.445\\
64.736912	0.455\\
64.764396	0.465\\
64.795236	0.475\\
64.824363	0.485\\
64.837745	0.495\\
64.84876	0.505\\
64.881666	0.515\\
64.941263	0.525\\
64.972902	0.535\\
65.030984	0.545\\
65.03438	0.555\\
65.053364	0.565\\
65.136942	0.575\\
65.172063	0.585\\
65.233927	0.595\\
65.286469	0.605\\
65.351056	0.615\\
65.37607	0.625\\
65.39241	0.635\\
65.39763	0.645\\
65.429351	0.655\\
65.438808	0.665\\
65.490752	0.675\\
65.503232	0.685\\
65.559656	0.695\\
65.600824	0.705\\
65.605989	0.715\\
65.618503	0.725\\
65.636754	0.735\\
65.671974	0.745\\
65.717614	0.76\\
65.725597	0.77\\
65.760308	0.78\\
65.780157	0.79\\
65.819072	0.8\\
65.850846	0.81\\
65.861316	0.82\\
65.884196	0.83\\
65.902047	0.84\\
65.959991	0.85\\
66.032954	0.86\\
66.081931	0.87\\
66.146703	0.88\\
66.3118	0.89\\
66.360203	0.9\\
66.45727	0.91\\
66.483874	0.92\\
66.544021	0.93\\
66.684609	0.94\\
66.701095	0.95\\
66.719055	0.96\\
66.752308	0.97\\
66.833333	0.98\\
66.864644	0.99\\
67.049498	1\\
};

\plotRinitOne{black}{forget plot, only marks}
   table[row sep=crcr]{%
54.938278	0\\
64.456802	0.255\\
64.837745	0.495\\
65.671974	0.745\\
67.049498	1\\
};



\node[align=center, font=\color{matGray}] at (0.3275,0.375) {\scriptsize non- };
\node[align=center, font=\color{matGray}] at (0.3275,0.3) {\scriptsize cooperative};
\node[align=center, font=\color{black}] at (25,0.7) {\scriptsize cooperative};

\draw [black, dashed, line width=1.0pt, rotate around={-16:(25,0.85)}] (25,0.85) ellipse [x radius=1.75, y radius=0.1];
\draw [matGray, dashed, line width=1.0pt, rotate around={-16:(0.3,0.15)}] (0.3,0.15) ellipse [x radius=1.75, y radius=0.1];
\end{axis}

\end{tikzpicture}%

%% file: 6_Conclusion.tex
\section{Conclusion}
\label{sec:conclusion}
We studied the localization performance of different estimation approaches for magneto-inductive \TA sensor nodes. In this context we analyzed the performance gain that can be obtained by enabling cooperation between the agent nodes, which are to be localized. Based on Cram\'er-Rao-Lower Bound analyses and numerical estimators, we found that having $10$ cooperating agents already reduces the position RMSE by a factor of $3$.

However, the agent cooperation also led to drastically increased complexity for the joint LS estimation problem and could not be solved within reasonable time by the Levenberg-Marquadt algorithm with multiple random initializations. We thus derived an analytical pairwise position estimator, which is based on the ML distance and direction estimates of a pair of three-axis coils. While this approach has worse localization accuracy, it is highly robust and its estimates are well-suited as initial value for the more complicated joint LS estimation. Combining both approaches enabled us to tightly approach the corresponding Cram\'er-Rao-Lower Bounds in all simulated cases. However, while this combined approach made it possible to actually realize the potential benefit of cooperating nodes, it still comes at the cost of longer computation times compared to \NC localization.

Overall, we see cooperative localization as an interesting extension to improve the localization accuracy for magneto-inductive networks. However, if its potential shall be fully exploited, the approach may be limited to applications with limited agent movement and no time-critical requirements.

%% file: AppendixA.tex
\appendices
\section{Derivatives of Channel Matrix}
\label{app:A_deriv}
In this appendix, the derivatives of the channel matrix $\bH \csub$ are given with respect to each element of the six-dimensional deployment vector ${\PSI _\m = [\p_\m \Tr,  \boldsymbol{\phi}_\m \Tr ]\Tr}$ of the agent $\m$.  Here,  the index $i=1,2,3$ denotes the scalar elements of either the spatial or the orientation part of the corresponding deployment vector. Similarly to \cite{bib:dumphart2017robust} and  \cite[Sec.~7.2]{bib:dumphart2020magneto}, the spatial derivatives are found as
\begin{align}
\gpsi{\,\bH \csub}{[\p _\m]_i} &= \O_\n\Tr \f{j \alpha \csub}{r^3\csub} \left ( \gpsi{\F}{ [\p _\m]_i}-\f{3 [\mathbf{u}\csub]_i \, \F}{r\csub} \right ) \O_\m \eqskip \\
\gpsi{\F}{[\p _\m]_i} &= \f{3}{2} \left ( \left ( \mathbf{u}\csub \gpsi{\mathbf{u}\csub\Tr}{[\p _\m]_i} \right )\Tr + \mathbf{u}\csub \gpsi{\mathbf{u}\csub\Tr}{[\p _\m]_i} \right) \eqskip \\
\gpsi{\mathbf{u}\csub\Tr}{[\p _\m]_i} &= \f{1}{r\csub} \left [  \mathbf{I}_3 -\mathbf{u}\csub \mathbf{u}\csub\Tr \right ] _{i,\ast} \eqskip
\end{align}
where $[.]_{i,\ast}$ denotes the $i$-th row of the corresponding matrix. 
The orientation derivatives follow as
\begin{align}
\gpsi{\,\bH \csub}{[\boldsymbol{\phi} _\m]_i} &= \O_\n\Tr \, \bHN \csub  \gpsi{\O _\m}{[\boldsymbol{\phi} _\m]_i}
\end{align}
with $\bHN \csub   =  \frac{j\coeff\csub}{r\csub^3} \left(\frac{3}{2}\mathbf{u}\csub\mathbf{u}\csub\Tr - \frac{1}{2}\mathbf{I}_3 \right)$ as in \Cref{subsec:pairwise_ML}.
Since $\O_\m$ is the result of multiplying the three rotation matrices of the individual Euler angles, its derivative with respect to a single angle is exemplarily found as $\gpsi{\O _\m}{[{\boldsymbol{\phi}} _\m]_1} = \gpsi{\O _{\alpha, \m} }{[{\boldsymbol{\phi}} _\m]_1}  \O _{\beta, \m} \O _{\gamma, \m} $ and the derivative of any of the Euler matrices only comprises the individual trigonometric derivations. 

%% file: ms.bbl
\begin{thebibliography}{10}
\providecommand{\url}[1]{#1}
\csname url@samestyle\endcsname
\providecommand{\newblock}{\relax}
\providecommand{\bibinfo}[2]{#2}
\providecommand{\BIBentrySTDinterwordspacing}{\spaceskip=0pt\relax}
\providecommand{\BIBentryALTinterwordstretchfactor}{4}
\providecommand{\BIBentryALTinterwordspacing}{\spaceskip=\fontdimen2\font plus
\BIBentryALTinterwordstretchfactor\fontdimen3\font minus
  \fontdimen4\font\relax}
\providecommand{\BIBforeignlanguage}[2]{{%
\expandafter\ifx\csname l@#1\endcsname\relax
\typeout{** WARNING: IEEEtran.bst: No hyphenation pattern has been}%
\typeout{** loaded for the language `#1'. Using the pattern for}%
\typeout{** the default language instead.}%
\else
\language=\csname l@#1\endcsname
\fi
#2}}
\providecommand{\BIBdecl}{\relax}
\BIBdecl

\bibitem{bib:abrudan2015distortion}
T.~E. Abrudan, Z.~Xiao, A.~Markham, and N.~Trigoni, ``Distortion rejecting
  magneto-inductive three-dimensional localization ({MagLoc}),'' \emph{IEEE
  Journal on Selected Areas in Communications}, vol.~33, no.~11, pp.
  2404--2417, 2015.

\bibitem{bib:akyildiz2015water}
I.~F. Akyildiz, P.~Wang, and Z.~Sun, ``Realizing underwater communication
  through magnetic induction,'' \emph{IEEE Communications Magazine}, vol.~53,
  no.~11, 2015.

\bibitem{bib:Sitti2015}
M.~Sitti, H.~Ceylan, W.~Hu, J.~Giltinan, M.~Turan, S.~Yim, and E.~Diller,
  ``Biomedical applications of untethered mobile milli/microrobots,''
  \emph{Proceedings of the IEEE}, vol. 103, no.~2, pp. 205--224, 2015.

\bibitem{bib:dumphart2020magneto}
G.~Dumphart, ``Magneto-inductive communication and localization: Fundamental
  limits with arbitrary node arrangements,'' Ph.D. dissertation, ETH Zurich,
  2020.

\bibitem{bib:Schlageter2001}
V.~Schlageter, P.-A. Besse, R.~Popovic, and P.~Kucera, ``Tracking system with
  five degrees of freedom using a {2D}-array of {Hall} sensors and a permanent
  magnet,'' \emph{Sensors and Actuators A: Physical}, vol.~92, no.~1, pp.
  37--42, 2001.

\bibitem{bib:SongHu2009}
S.~Song, C.~Hu, M.~Li, W.~Yang, and M.~Q.-H. Meng, ``Real time algorithm for
  magnet's localization in capsule endoscope,'' in \emph{IEEE International
  Conference on Automation and Logistics}, 2009.

\bibitem{bib:dumphart2017robust}
G.~Dumphart, E.~Slottke, and A.~Wittneben, ``{Robust Near-Field 3D Localization
  of an Unaligned Single-Coil Agent Using Unobtrusive Anchors},'' in \emph{IEEE
  PIMRC}, 2017.

\bibitem{bib:Hu2005}
C.~Hu, M.~Q.-H. Meng, and M.~Mandal, ``Efficient magnetic localization and
  orientation technique for capsule endoscopy,'' \emph{International Journal of
  Information Acquisition}, vol.~2, no.~01, pp. 23--36, 2005.

\bibitem{bib:XieTMC2015}
H.~Xie, T.~Gu, X.~Tao, H.~Ye, and J.~Lu, ``A reliability-augmented particle
  filter for magnetic fingerprinting based indoor localization on smartphone,''
  \emph{IEEE Transactions on Mobile Computing}, vol.~15, no.~8, pp. 1877--1892,
  2015.

\bibitem{bib:KyprisTGRS2016}
O.~Kypris, T.~E. Abrudan, and A.~Markham, ``Magnetic induction-based
  positioning in distorted environments,'' \emph{IEEE Transactions on
  Geoscience and Remote Sensing}, vol.~54, no.~8, pp. 4605--4612, 2016.

\bibitem{bib:dumphart2019practical}
G.~Dumphart, H.~Schulten, B.~Bhatia, C.~Sulser, and A.~Wittneben, ``Practical
  accuracy limits of radiation-aware magneto-inductive {3D} localization,'' in
  \emph{2019 IEEE International Conference on Communications (ICC) Workshops},
  2019.

\bibitem{bib:shen2010fundamental}
Y.~Shen, H.~Wymeersch, and M.~Z. Win, ``Fundamental limits of wideband
  localization — part {II}: Cooperative networks,'' \emph{IEEE Transactions
  on Information Theory}, vol.~56, no.~10, pp. 4981--5000, 2010.

\bibitem{bib:WinCM2011}
M.~Z. Win, A.~Conti, S.~Mazuelas, Y.~Shen, W.~M. Gifford, D.~Dardari, and
  M.~Chiani, ``Network localization and navigation via cooperation,''
  \emph{IEEE Communications Magazine}, vol.~49, no.~5, 2011.

\bibitem{bib:buehrer2018collaborative}
R.~M. Buehrer, H.~Wymeersch, and R.~M. Vaghefi, ``Collaborative sensor network
  localization: Algorithms and practical issues,'' \emph{Proceedings of the
  IEEE}, vol. 106, no.~6, pp. 1089--1114, 2018.

\bibitem{bib:dumphart2016stochastic}
G.~Dumphart and A.~Wittneben, ``Stochastic misalignment model for
  magneto-inductive {SISO} and {MIMO} links,'' in \emph{IEEE PIMRC}, 2016.

\bibitem{bib:mirsky1975traceNeumann}
L.~Mirsky, ``A trace inequality of {John von Neumann},'' \emph{Monatshefte
  f{\"u}r mathematik}, vol.~79, no.~4, pp. 303--306, 1975.

\bibitem{bib:schonemann1966generalizedProcrustes}
P.~H. Sch{\"o}nemann, ``A generalized solution of the orthogonal {Procrustes}
  problem,'' \emph{Psychometrika}, vol.~31, no.~1, pp. 1--10, 1966.

\bibitem{bib:kabsch1976solution}
W.~Kabsch, ``A solution for the best rotation to relate two sets of vectors,''
  \emph{Acta Crystallographica Section A: Crystal Physics, Diffraction,
  Theoretical and General Crystallography}, vol.~32, no.~5, pp. 922--923, 1976.

\bibitem{bib:levenberg1944method}
K.~Levenberg, ``A method for the solution of certain non-linear problems in
  least squares,'' \emph{Quarterly of applied mathematics}, vol.~2, no.~2, pp.
  164--168, 1944.

\bibitem{bib:lsqnonlin}
\BIBentryALTinterwordspacing
{MathWorks Inc.} (2020) Documentation: lsqnonlin - solve nonlinear
  least-squares (nonlinear data-fitting) problems. [Online]. Available:
  \url{https://de.mathworks.com/help/optim/ug/lsqnonlin.html}
\BIBentrySTDinterwordspacing

\end{thebibliography}
